\documentclass[12pt,notitlepage,a4paper]{article}

\pdfoutput=1

\usepackage{a4wide}
\usepackage{epsfig}
\usepackage{color,graphicx}
\usepackage{cite}
\usepackage{amssymb,amsmath}

\newcommand{\be}{\begin{equation}}
\newcommand{\ee}{\end{equation}}
\newcommand{\bea}{\begin{eqnarray}}
\newcommand{\eea}{\end{eqnarray}}

\begin{document}

\begin{center}  

\vskip 2cm 

\centerline{\Large {\bf On the torus compactifications of $Z_2$ orbifolds of E-string theories}}

\vskip 1cm

\renewcommand{\thefootnote}{\fnsymbol{footnote}}

   \centerline{
    {\large \bf Gabi Zafrir${}^{a}$} \footnote{gabi.zafrir@ipmu.jp}}

\vspace{1cm}
\centerline{{\it ${}^a$ Kavli IPMU (WPI), UTIAS, the University of Tokyo, Kashiwa, Chiba 277-8583, Japan}}
\vspace{1cm}

\end{center}

\vskip 0.3 cm

\setcounter{footnote}{0}
\renewcommand{\thefootnote}{\arabic{footnote}}   
   
\begin{abstract}

We consider the torus compactifications with flux of a class of $6d$ $(1,0)$ SCFTs that can be engineered as the low-energy theories on M$5$-branes near an M$9$-plane on a $C^2/Z_2$ singularity. Specifically, we concentrate on the two SCFTs where the $Z_2$ orbifold action acts non-trivially on the $E_8$ global symmetry. We analyze this problem by compactifying to $5d$, where we can exploit the relation to $5d$ duality domain walls. By a suitable guess of the domain wall theories, the resulting $4d$ theories can be conjectured. These can then be tested by comparing their properties, notably anomalies and symmetries, against the $6d$ expectations. These constructions lead to various interesting $4d$ phenomena like dualities and symmetry enhancements.

\end{abstract}
 
 \newpage
 
\tableofcontents

\section{Introduction}

A long standing problem in quantum field theory is the study of the dynamics of gauge theories. Even concentrating only on the subset of $4d$ supersymmetric gauge theories, we already have a wide variety of interesting strong coupling phenomena like duality and symmetry enhancement. This leads to the desire for some organizing principle allowing one to motivate and predict the occurrences of such phenomena. One such principle, is the realization of the $4d$ theory through the compactification of a $6d$ SCFT. Initiated in \cite{Gai} for $4d$ $\mathcal{N}=2$ SCFT, it has proven to be an effective tools for constructing SCFTs and dualities between them.

Building on this work, there have been many attempts to generalize this in multiple directions, which are in fact so numerous that we will not attempt to review them all here. Instead we concentrate on the generalizations, retaining most of the features of the original construction, but leading to $4d$ $\mathcal{N}=1$ SCFTs. This can be achieved by starting from some $6d$ $\mathcal{N}=(1,0)$ SCFT and reducing to $4d$ through a compactification preserving $4d$ $\mathcal{N}=1$ SUSY.

There are by now various examples of this, where the chosen $(1,0)$ SCFT is the $(2,0)$ theory\footnote{This $6d$ SCFT of course has $\mathcal{N}=(2,0)$ SUSY, but still one can consider compactifications of it that only preserve $4d$ $\mathcal{N}=1$. In this type of constructions, the $(2,0)$ theory is not vastly different from any other $(1,0)$ SCFT.}\cite{BTW,BBBW} or its orbifold cousins\cite{GR,HM,FHU,RVZ,BHMRTZ,KRVZ,KRVZ1,KRVZ2}, where the references given are by no means an exhaustive list. Recently examples were also given for some classes of the so called minimal $6d$ SCFTs\cite{RZ}.

Starting in \cite{KRVZ} and further developed in \cite{KRVZ1,KRVZ2}, a method for studying the torus compactification of in principle generic $6d$ SCFTs, with fluxes in their global symmetry, was proposed and studied for the cases of orbifolds of the $A$ type $(2,0)$ theory. This method maps the problem to the study of $5d$ domain walls between different low-energy $5d$ gauge theories all having the $6d$ SCFT as their UV completion. Unfortunately, a systematic understanding of these $5d$ domain walls is still lacking, and as a result, this method cannot be used to systematically study compactifications. However, progress can still be made by studying specific examples of $5d$ domain walls, which are determined either from insight or by guesswork, generally by operating under the assumption that these are sufficiently simple. This was used to study various cases in \cite{KRVZ1,KRVZ2}.

The purpose of this article is to further use this method to try and study additional cases. A natural question then is why should we bother in undertaking this. Our motivations for this are twofold. First, it is interesting to see how far one can get using the method of $5d$ domain walls. In the cases studied, these methods were sufficient to construct a large class of compactification. However, this may not be a generic feature. As we lack a systematic understanding of this process, it seems necessary to analyze examples to get a better understanding of what can be expected from these methods.

The second motivation is more rooted in studying the dynamics of $4d$ theories. One desire in the study of compactifications to $4d$ is to uncover interesting dynamical phenomena in $4d$, like enhancement of symmetry and duality. For the former, we are especially interested in cases with exceptional symmetries that are not easy to generate with Lagrangian theories. As for dualities, these tend to be ubiquitous in these constructions as there are usually many ways to build the same compactification. Nevertheless, in most of the cases studied so far the dualities encountered were mostly known. However, there are cases where one gets dualities, which are more interesting, and don't quite appear to reduce to known ones, at least at first sight. This occurs when there are at least two different low-energy $5d$ gauge theories having the $6d$ SCFT as their UV completion, a phenomenon refereed to as a $5d$ duality. In these cases it may be possible to engineer the same $4d$ theory, but with domain walls that extrapolate between the different cases. The resulting dualities then can be seen as a $4d$ manifestation of the $5d$ duality.

 An interesting family of SCFTs to study then is the ones describable as low-energy theories on M$5$-branes probing a $C^2/Z_k$ singularity in the presence of an M$9$-plane. These theories then can inherit in some cases exceptional symmetries from their parent theories, and further are known to posses multiple $5d$ gauge theory descriptions\cite{Zaf2,HKLY}. Here we shall only concentrate on the case of $k=2$ and with a non-trivial action of the $Z_2$ on the $E_8$ global symmetry, reserving other cases for future study.

The structure of this article is as follows. We begin in section $2$ by discussing several aspects of the $6d$ SCFTs that are of interest to us here. Specifically, we touch upon their realization, notable operator spectrum and anomalies. We also consider their reduction to $5d$, which will be useful later. In section $3$ we consider the $4d$ theories. Specifically, using a particular $5d$ gauge theory description we conjecture the $4d$ theories resulting from torus compactifications of the studied $6d$ SCFTs with specific fluxes in their global symmetries. We then test this conjecture by various methods, notably by comparing anomalies. The results of this section suggest various interesting cases of symmetry enhancement. We continue our study of the $4d$ theories in section $4$, where we conjecture another class of $4d$ theories with an expected $6d$ origin, now using a different domain wall. These theories turn out to be only mass deformation of a direct $6d$ compactification, a claim we support with various pieces of evidence. Nevertheless, this still leads to interesting proposals for symmetry enhancements and 　duality. We end in section $5$ with some conclusions.  

\section{$6d$ and $5d$ discussion}

Here we consider the $6d$ SCFTs described by M5-branes, near an M9-plane, probing a $C^2/Z_2$ singularity. There are in fact three different $6d$ SCFTs associated with this setup, differing by how the $Z_2$ acts on the $E_8$ symmetry\cite{HMRV}. Besides the trivial action, preserving the $E_8$, there are two non-trivial ones breaking $E_8$ to either its $SO(16)$ or $SU(2) \times E_7$ maximal subgroups. Here we will be concerned only with the two SCFTs associated with the non-trivial action.

First we consider some properties of the SCFTs, starting with their global symmetries. Both SCFTs have the symmetry preserved by the $Z_2$ embedding inside $E_8$, which as mentioned is either $SO(16)$ or $SU(2) \times E_7$ depending on the choice of embedding. In addition they have an $SU(2)$ coming from the orbifold. Finally we have the $SU(2)\times SU(2)$ isometry of $C^2/Z_2$ where one $SU(2)$ is the R-symmetry and the other is a global symmetry. So we conclude that the SCFTs have the global symmetry of $SU(2) \times SU(2)\times SO(16)$ for one embedding and $SU(2)\times SU(2) \times SU(2) \times E_7$ for the other.

\begin{figure}
\center
\includegraphics[width=0.85\textwidth]{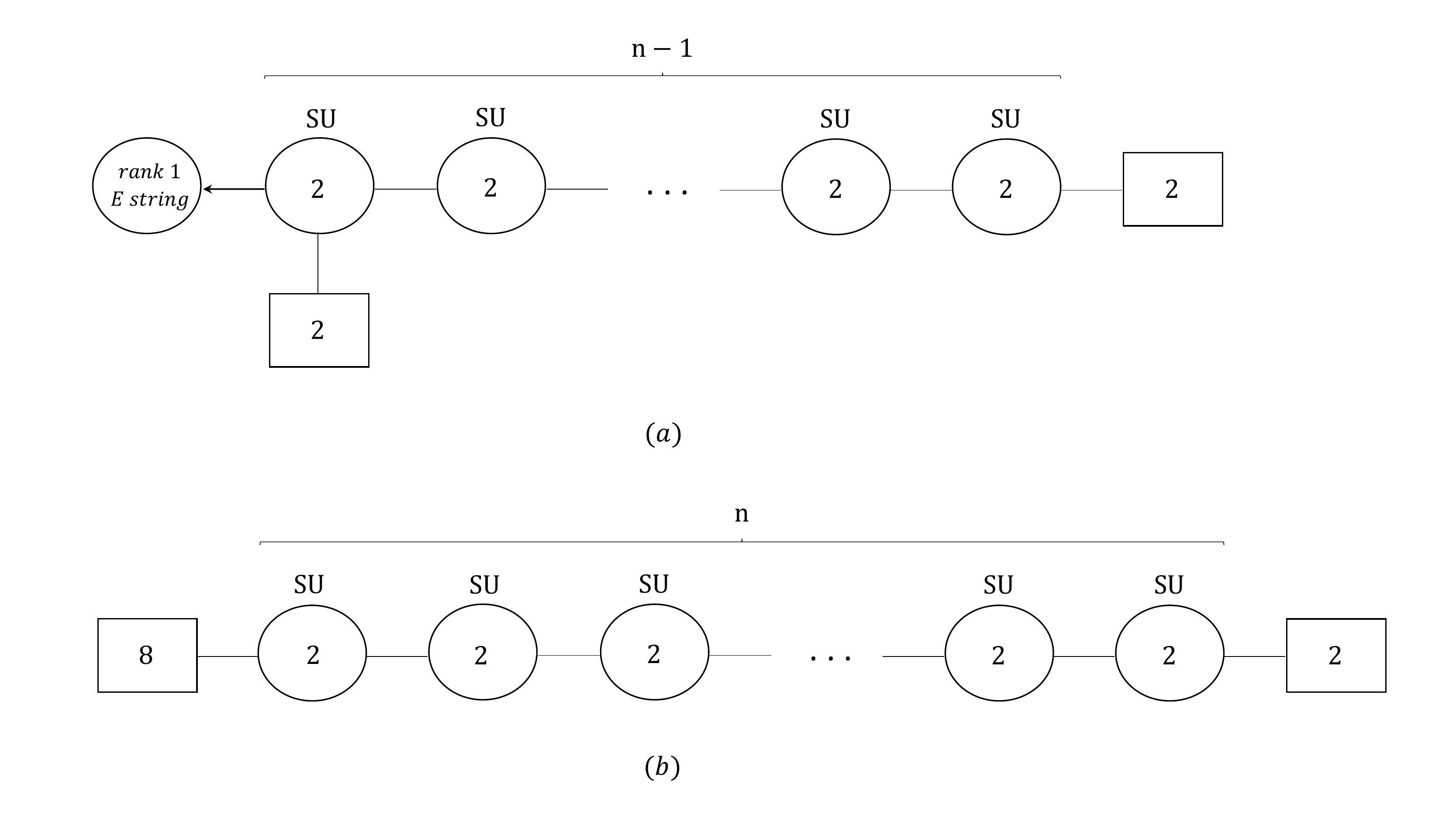} 
\caption{The low-energy description on the tensor branch of the $6d$ SCFTs associated with $Z_2$ orbifolds of E-string. Quiver (a) describes the case of $SU(2) \times E_7$ preserving embeddings while quiver (b) the case of $SO(16)$ preserving embeddings. In quiver (a) the arrow into the rank $1$ E-string theory represents gauging an $SU(2)\subset E_8$ of the global symmetry of the E-string theory.}
\label{6dQuivers}
\end{figure}

To make further progress it is convenient to employ the low-energy description of the SCFTs on a generic point in the tensor branch. Then the SCFT associated with $SO(16)$ has a gauge theory description while the one associated with $SU(2) \times E_7$ as a description as a gauge theory connected to the rank $1$ E-string theory via gauging. The two descriptions are shown in figure \ref{6dQuivers}. We next employ these to uncover more properties of these SCFTs.

First, it is instructive to study the global symmetry also from the quiver low-energy description. The part associated with the commutant of the $Z_2$ embedding in $E_8$ is given by the 'matter' on the left side of the quiver. Specifically, for $SO(16)$ it is the symmetry rotating the $8$ flavors. For $SU(2) \times E_7$, the $E_7$ is the commutant of $SU(2)$ in $E_8$, and the $SU(2)$ is associated with rotating the two flavors for the leftmost $SU(2)$ gauge group. In both quivers we have an $SU(2)$ associated with the two flavors of the rightmost $SU(2)$ gauge group. This is identified with the $SU(2)$ associated with the singularity.

The symmetry associated with rotating two flavors of an $SU(2)$ gauge group is $SO(4) = SU(2) \times SU(2)$, while $SU(2) \times SU(2)$ bifundamentals also have $SU(2)$ symmetries associated with them. Thus, we naively reach a contradiction where the gauge theory description suggests we have an order $n$ additional $SU(2)$ global symmetries. However, it is argued that this is incorrect and that the additional $SU(2)$'s are accidental enhancements\cite{MOTZ}. This hangs on the observation of \cite{OSTY2} that for $SU(2)+4F$ the symmetry of the SCFT appears to be $SO(7)$ and not $SO(8)$, a claim which was motivated by studying the compactifications of this SCFT to $5d$ and $4d$. In the quivers in figure \ref{6dQuivers}, each node is an $SU(2)+4F$ and the multitude of $SU(2)$ global symmetries come from the $SU(2)^4$ subgroup of $SO(8)$. However, $SO(7)$ only has an $SU(2) \times SU(2)^2$ maximal subgroup, where an $SU(2)^2$ in $SO(8)$ is reduced to the diagonal $SU(2)$. Thus, one concludes that at the SCFT there is only one $SU(2)$ which is the diagonal combination of all bifundamental $SU(2)$'s and one of the $SU(2)$'s rotating the two flavors at the end\footnote{This can also be reached by continuity from the case where the groups are $SU(k)$. Then for $k>2$ the $SU(2)$ global symmetry is broken to $U(1)$ which suffers from gauge anomalies leaving only the diagonal combination non-anomalous.}. We identify this $SU(2)$ with the isometry of $C^2/Z_2$. 

Having mapped the symmetries to the quivers, we next study the matter spectrum in these theories. We shall begin with the $SO(16)$ preserving embedding. First we have the conserved currents for the global symmetries. These contain scalar operators in the adjoint of the global symmetry which are the analogue of the moment map operators in $4d$. These are given by various mesons in the quiver and are in the $\bold{3}$ of $SU(2)_R$.

We next have the gauge invariant made from all bifundamentals and the flavors at the two ends. This gives a Higgs branch generator in the $\bold{n+2}$ of $SU(2)_R$ that is in the $(\bold{2},\bold{n+1},\bold{16})$ of $SU(2) \times SU(2)\times SO(16)$\footnote{We shall adopt the notation where the $SU(2)$ associated with the singularity is written first and symmetries associated with $E_8$ last.}. This concludes the states we can observe perturbatively. Additionally we have a non-perturbative contribution from the instanton strings of the leftmost $SU(2)$. Essentially, we can consider taking the infinite coupling limit for the leftmost $SU(2)$, while keeping the other couplings finite. Then this $SU(2)$ gauge theory is expected to be completed to a $6d$ SCFT known as the $(D_5,D_5)$ minimal conformal matter\cite{ZHTV}. This SCFT is known to have a Higgs branch generator in the the $\bold{4}$ of $SU(2)_R$ and in the $\bold{512}$ of its $SO(20)$ global symmetry, that arises non-perturbatively from the gauge theory viewpoint\cite{HMa,KRVZ1}. Such an operator should then also arise in the class of theories we consider here, but as $SU(2) \subset SO(20)$ here is gauged, we need to project it to the gauge invariant contribution. Therefore, it is expected to contribute a Higgs branch generator in the $\bold{4}$ of $SU(2)_R$ and the $(\bold{1},\bold{2},\bold{128})$ of $SU(2) \times SU(2)\times SO(16)$. 

From this we can attempt to identity the global structure of the symmetry. It appears to be $\frac{(SU(2) \times SU(2)\times Spin(16))}{Z_2 \times Z_2}$, where one $Z_2$ is the combination of the center of the orbifold $SU(2)$ and the center of $Spin(16)$ with non-trivial action on the vector and the $\bold{128'}$ spinor. The other $Z_2$ is the combination of the center of the isometry $SU(2)$ and the center of $Spin(16)$ with non-trivial action on the vector and the $\bold{128}$ spinor when $n$ is odd, while for even $n$ it also involves the center of the orbifold $SU(2)$. 

Next we move to the $SU(2) \times E_7$ preserving embedding. We still have the conserved currents for the global symmetries with their associated scalar operators. Besides these there is also the Higgs branch generator given by the gauge invariant combination of all the bifundamentals and the flavors at the two ends. It is in the $\bold{n+1}$ of $SU(2)_R$ and in the $(\bold{2},\bold{n+1},\bold{2},\bold{1})$ of $SU(2) \times SU(2)\times SU(2)\times E_7$. Next we move to the non-perturbative sector. The rank $1$ E-string has an $E_8$ global symmetry which is broken by gauging. Yet it still retains the scalar operators associated with the conserved currents of $E_8$. These include ones in the $\bold{56}$ of $E_7$ and the doublet of the $SU(2)$ gauge symmetry. While by themselves gauge variant, we can combine them with matter fields to build a gauge invariant. The simplest combination involves just using the flavors of the leftmost $SU(2)$ gauge group. This gives a Higgs branch generator in the $\bold{4}$ of $SU(2)_R$ that is in the $(\bold{1},\bold{2},\bold{2},\bold{56})$ of $SU(2) \times SU(2)\times SU(2)\times E_7$. Alternatively we can use the bifundamentals and flavors for the rightmost $SU(2)$ gauge group to build an invariant. This gives an operator in the $\bold{n+2}$ of $SU(2)_R$ that is in the $(\bold{2},\bold{n},\bold{1},\bold{56})$.  

From this we can attempt to identity the global structure of the symmetry. It appears to be $\frac{(SU(2) \times SU(2)\times SU(2)\times E_7)}{Z_2 \times Z_2}$, where one $Z_2$ is the combination of the center of the orbifold $SU(2)$, the center of the embedding $SU(2)$ and the center of $E_7$. The other $Z_2$ is the combination of the center of the isometry $SU(2)$ and the center of the embedding $SU(2)$ when $n$ is odd, while for even $n$ it also involves the center of the orbifold $SU(2)$.

\subsection{Special cases}

For small value of $n$ the symmetry enhances and these cases have some special features. First in the $n=1$ case for the $SO(16)$ preserving embedding, the $SU(2) \times SU(2)\times SO(16)$ enhances to $SO(20)$ and the theory reduces to the minimal $(D_5,D_5)$ conformal matter. Torus compactifications for these cases were discussed in \cite{KRVZ1,KRVZ2}, and the higher $n$ cases can be thought of as generalizations of these cases.

In the $SU(2) \times E_7$ preserving embedding, the interesting special case is when $n=2$. Then the three $SU(2)$ groups enhance to form the group $SO(7)$. This SCFT is known as the $SO(7) \times E_7$ conformal matter\cite{ZHTV}. In this case the previously mentioned states merge to form just two basic states. One are just the moment map operators associated with the global symmetry. The second is a Higgs branch generator in the $\bold{4}$ of $SU(2)_R$ and the $(\bold{8},\bold{56})$ of $SO(7) \times E_7$. This leads us to conclude that the global symmetry is globally $\frac{Spin(7) \times E_7}{Z_2}$ where the $Z_2$ is the diagonal center.
 
\subsection{Reduction to lower dimensions}

We next consider the reduction of these theories on circles to lower dimensions without flux. This was studied in \cite{OSTY1,DVX,Zaf2,HKLY,OS,MOTZ}. Generically, when we reduce $6d$ theories on a circle we can incorporate an holonomy in the global symmetry, and different holonomies can lead to different $5d$ theories. When reduced to $5d$, with a specific choice of holonomy, the two theories have an effective description as $SU(N)_0+2AS+8F$ gauge theories, where $N=2n$ for the $SU(2) \times E_7$ preserving embedding and $N=2n+1$ for the $SO(16)$ preserving embedding\footnote{Here we adopt an abbreviated notation for writing gauge theories with $8$ supercharges. The groups written are the gauge groups, and subscript denotes Chern-Simons level when applicable. Multiple groups are written with $\times$, where it is understood that there is a bifundamental hypermultiplet between each adjacent group. Matter content for the groups are denotes by $+$ with a letter providing the representation under the gauge group and a number for the number of hypermultiplets. We use $F$ for the fundamental representation and $AS$ for the antisymmetric representation.}. Another useful $5d$ description, given by a different holonomy choice, is as a $USp\times USp$ quiver gauge theory. Specifically, the $SO(16)$ preserving embedding has a $4F+USp(2n)\times USp(2n)+4F$ description while the $SU(2) \times E_7$ preserving embedding has a $6F+USp(2n)\times USp(2n-2)+2F$ description. These $5d$ descriptions will play an important role later. There can be other descriptions related to other holonomy choices though we won't carry a complete listing here. The $5d$ descriptions can be employed to study the spectrum of the reduced theory though we won't pursue this here.

The special cases materialize in the $5d$ $SU(N)$ description as symmetry enhancements due to special properties of the antisymmetric representation for small $N$. Particularly when $N=3$, the antisymmetric is the same as the antifundametal and the symmetry combines to one big group. For $N=4$, corresponding to the $SO(7) \times E_7$ conformal matter case, the antisymmetric representation becomes real and the symmetry rotating the antisymmetrics enhances. 

We can further reduce to $4d$ where these reduce to class S theories associated with an A type $(2,0)$ theory on a $3$ punctured sphere. We can also reduce to $3d$ where there is a mirror dual. Both of these can be employed to study the spectrum of the reduced $6d$ SCFT. The $3d$ mirror is especially convenient and most of the operators we identified can be seen from there too.

\subsection{Tubes and $5d$ gauge theory descriptions}

The $5d$ gauge theory descriptions are particularly useful for conjecturing the $4d$ theories resulting from compactifications of the $6d$ SCFTs on a torus with flux. This was first pointed out in \cite{KRVZ} an later expanded upon in \cite{KRVZ1,KRVZ2}. We shall next review the general idea, referring to these references for more details. 

Consider the reduction of the $6d$ SCFT on a circle to $5d$. As previously mentioned, in this reduction we have a choice of holonomy in the $6d$ global symmetry. Consider then the following choice. We take the holonomy to be dependent on a space direction in the $5d$ spacetime, say $x^5$, and have the profile of a step function. Specifically, at $x^5>0$ we have it be the holonomy leading to a particular $5d$ gauge theory description, and at $x^5<0$ we have it be the holonomy leading to another $5d$ gauge theory description. Note, that the two gauge theory descriptions can be the same, but associated with two different holonomies leading to the same $5d$ gauge theory description.     

In $5d$ this leads to a duality domain wall, which is a codimension $1$ defect, here located at $x^5=0$, which extrapolates between two dual $5d$ gauge theories, that is two $5d$ gauge theory descriptions of the same underlying SCFT. Such duality domain wall defect were studied in \cite{GC}. There it was found that these can preserve half the supersymmetry leading to $\mathcal{N}=1$ SUSY on the $4d$ theory on the domain wall. As usual in such defects, in general there is a $4d$ theory, potentially strongly coupled, living on the domain wall.

In $6d$, however, we have a variable holonomy which leads to a Delta function like curvature on $x^5=0$. This leads to flux in the $6d$ global symmetry on the surface spanned by $x^5$ and the compactification circle. Now consider further compactifying on $x^5$. The holonomies at the two ends must be compatible for this to be possible, and we shall assume this is the case. In the $6d$ perspective, we have a compactification of the $6d$ SCFT on a torus with flux, which is what we are considering. 

On the $5d$ side, we have a collection of $5d$ gauge theories connected via duality domain walls, where here we have generalized to the case where the holonomy may have multiple steps. It is now straightforward to reduce this system to $4d$. The $5d$ gauge theories are IR free, and so should just reduce to the analogous $4d$ gauge theories. These are then coupled to one another via the theory living on the domain walls. Thus, we see that understanding compactifications of $6d$ SCFTs on tori with fluxes is related to the study of $4d$ $\mathcal{N}=1$ theories living on $5d$ duality domain walls.  

Instead of considering compactificatons on tori, it is convenient to consider compactifications on tubes, that is a sphere with two punctures. The torus cases can be generated from the tubes by gluing the two ends as we shall describe momentarily. For this case, we take $x^5$ to be an interval. The punctures are manifested on the $5d$ system as half-BPS boundary conditions at the ends of the interval.

Here we shall only consider one type of punctures, generalizing the maximal punctures of class S. In terms of the $4d$ fields living on the boundary of the interval, these are given by Dirichlet boundary conditions for the $4d$ $\mathcal{N}=1$ vector multiplet in the $\mathcal{N}=2$ vector multiplet, and one of the chiral fields in the hypermultiplets while giving Nuemann boundary conditions for the adjoint chiral in the $\mathcal{N}=2$ vector multiplet, and the remaining chiral fields in the hypermultiplets. 

For example consider the case where the $5d$ gauge theory description is $SU(N)_0+2AS+8F$. Then on the boundary we can decompose the $5d$ fields into $4d$ $\mathcal{N}=1$ multiplets. In this case the maximal puncture is defined by Dirichlet boundary conditions for the $SU(N)$ $\mathcal{N}=1$ vector multiplet, the two chiral multiplets in the conjugate antisymmetric representation of $SU(N)$, and the eight chiral multiplets in the antifundamental representation of $SU(N)$. The remaining fields, the $SU(N)$ adjoint chiral, the two chiral multiplets in the antisymmetric representation of $SU(N)$, and the eight chiral multiplets in the fundamental representation of $SU(N)$, are given Nuemann boundary conditions. Of course we could have chosen to give Dirichlet boundary conditions to the fundamentals and Nuemann boundary conditions for the antifundamentals, and similarly for the fields in the antisymmetric hypers. These different choices define equivalent though slightly different punctures, and we shall refer to these different choices as the color of the puncture\footnote{More generally the notion of a color of a puncture refers to the charges of the operators associated with the puncture under the $6d$ global symmetry. The chiral fields which are given Nuemann boundary conditions in the hypermultiplets generically provide such operators, and as their charges under the $6d$ global symmetry changes depending on which of them is chosen, this choice naturally leads to different colors. However, there may be other colors not directly related to these choices as we shall see in the next section.}. This generalizes the notion of color from $\mathcal{N}=1$ class S and class $S_k$\cite{BG,ABMS,GR,RVZ}. Here we shall be mostly concerned with two types of colors, one as defined above and the other with the Dirichlet and Nuemann boundary conditions for the hypermultiplets reversed. 

There are two important properties about the punctures that we can extract from this construction. One is the global symmetry associated with the punctures. In general as we have given Dirichlet boundary conditions for the $\mathcal{N}=1$ vector multiplet, the $5d$ gauge symmetry should become non-dynamical and so is a global symmetry associated with the puncture. For the cases at hand, we expect at least two types of maximal punctures with different global symmetries corresponding to the two different $5d$ gauge theory descriptions.   

We can also consider the process of gluing two punctures of the same type. Consider for instance taking two tubes each ending with a puncture and glue them along the punctures to create a single tube. Reducing again to $5d$ we now have two intervals, with a given boundary condition, that are joined to form a line. As the boundary conditions projects out the fields that were given Dirichlet boundary conditions, when we do this gluing, these have to be reintroduced by hand. Therefore, the process of gluing is done by connecting the two tubes via the fields given Dirichlet boundary conditions. Since we give Dirichlet boundary conditions for the $\mathcal{N}=1$ vector multiplet, the gluing involves gauging the puncture symmetry with an $\mathcal{N}=1$ vector multiplet. Additionally, we must also add chiral fields, $\Phi_i$, for every chiral field, in the hypermultiplet, with Dirichlet boundary conditions. These should be coupled to the analogous fields given Nuemann boundary conditions for both punctures, let's call them $M_i$ and $\tilde{M}_i$ respectively, via the superpotential $W = M_i \Phi_i - \tilde{M}_i \Phi_i$. The rational behind this is that this forces $M_i=\tilde{M}_i$ in the chiral ring which is exactly what we need. Gluing punctures of the same color does not break any of the $6d$ global symmetry, as $M_i$ and $\tilde{M}_i$ have the same charges under them, which are by construction opposite to those of $\Phi_i$. However, when gluing punctures of differing colors, some of the $6d$ global symmetry is broken by the superpotential.     

Finally we wish to discuss how this is used in this article to determine the 4d theories. Here the $6d$ SCFTs enjoy two different $5d$ descriptions, one as an $SU(N)_0+2AS+8F$ gauge theory and another as a $USp\times USp$ quiver theory. We can then consider various domain walls extrapolating between any combination of these. When reduced to $4d$ these should give a $4d$ theory having the $5d$ descriptions on both sides, connected via the fields living on the domain walls. Here we shall assume that these fields are just some collection of free fields, specifically, these will be guessed to be bifundamentals between the gauge groups on the two sides potentially with an additional chiral fields flipping baryons made of these\footnote{The process of flipping an operator $O$ is defined as adding a chiral field, $F_O$, and coupling it to $O$ via the superpotential $W= O F_O$. This has the effect of eliminating $O$ from the chiral ring, replacing it with $F_O$.}. These conjectured theories are then tested by various methods that we shall explain in greater detail later. One such test that is quite useful will be comparing the anomalies of these $4d$ theories from those expected from the anomalies of the $6d$ theories, as encoded in the anomaly polynomial. Next, we shall present the anomaly polynomial and later use it to derive the expected anomalies of the $4d$ theories.

\subsection{Anomaly polynomial}

We can use the low-energy gauge theory description on the tensor branch to calculate the anomaly polynomial of these $6d$ SCFTs using the methods of \cite{OSTY}. For the $SO(16)$ preserving embedding we find:

\bea
I_{SO(16)} & = & \frac{n(8n^2 + 12n + 3)}{12} C^2_2(R) - \frac{n(6n+7)}{24} p_1 (T) C_2(R) + \frac{3n+1}{24} p_1 (T) C_2(SO(16))_{\bold{16}} \\ \nonumber & + & \frac{1}{3} p_1 (T) C_2(SU(2)_E)_{\bold{2}} + \frac{n(3n+1)}{12} p_1 (T) C_2(SU(2)_F)_{\bold{2}} - \frac{n(n+1)}{2} C_2(R) C_2(SO(16))_{\bold{16}} \\ \nonumber & - & 2n C_2(R) C_2(SU(2)_E)_{\bold{2}} - \frac{n(n+1)(4n-1)}{3} C_2(R) C_2(SU(2)_F)_{\bold{2}} + \frac{(3n+2)}{24} C^2_2(SO(16))_{\bold{16}} \\ \nonumber & + & \frac{1}{2} C_2(SU(2)_E)_{\bold{2}} C_2(SO(16))_{\bold{16}} + \frac{n^2}{2} C_2(SU(2)_F)_{\bold{2}} C_2(SO(16))_{\bold{16}} + \frac{2}{3} C^2_2(SU(2)_E)_{\bold{2}} \\ \nonumber & + & 2n C_2(SU(2)_E)_{\bold{2}} C_2(SU(2)_F)_{\bold{2}} + \frac{2 n^3}{3} C^2_2(SU(2)_F)_{\bold{2}} - \frac{1}{6} C_4(SO(16))_{\bold{16}} \\ \nonumber & + & \frac{(15n+8)( 7 p^2_1 (T) - 4 p_2(T))}{2880} .
\eea 

Here we adopted the conventions of \cite{RVZ}. We also use $SU(2)_E$ and $SU(2)_F$ for the $SU(2)$ global symmetries associated with the singularity and the isometry of $C^2/Z_2$ respectively. For the $SU(2) \times E_7$ preserving embedding we find:

\bea
I_{SU(2) \times E_7} & = & \frac{(8n^3 + 50n - 45)}{24} C^2_2(R) - \frac{(12n^2+2n-3)}{48} p_1 (T) C_2(R) + \frac{n}{48} p_1 (T) C_2(E_7)_{\bold{56}} \\ \nonumber & + & \frac{1}{3} p_1 (T) C_2(SU(2)_E)_{\bold{2}} + \frac{(3n-2)}{12} p_1 (T) C_2(SU(2)_I)_{\bold{2}} + \frac{n(3n-2)}{12} p_1 (T) C_2(SU(2)_F)_{\bold{2}} \\ \nonumber & - & \frac{n^2}{12} C_2(R) C_2(E_7)_{\bold{56}} - (2n-1) C_2(R) C_2(SU(2)_E)_{\bold{2}} - (n^2-1) C_2(R) C_2(SU(2)_I)_{\bold{2}} \\ \nonumber & - & \frac{n(n-1)(4n+1)}{3} C_2(R) C_2(SU(2)_F)_{\bold{2}} + \frac{n}{288} C^2_2(E_7)_{\bold{56}} + \frac{1}{12} C_2(SU(2)_E)_{\bold{2}} C_2(E_7)_{\bold{56}} \\ \nonumber & + & \frac{(n-1)}{12} C_2(SU(2)_I)_{\bold{2}} C_2(E_7)_{\bold{56}} + \frac{n(n-1)}{12} C_2(SU(2)_F)_{\bold{2}} C_2(E_7)_{\bold{56}} + \frac{2}{3} C^2_2(SU(2)_E)_{\bold{2}} \\ \nonumber & + & (2n-1) C_2(SU(2)_E)_{\bold{2}} C_2(SU(2)_F)_{\bold{2}} + (n^2-n+1) C_2(SU(2)_I)_{\bold{2}} C_2(SU(2)_F)_{\bold{2}} \\ \nonumber & + & C_2(SU(2)_E)_{\bold{2}} C_2(SU(2)_I)_{\bold{2}} + \frac{(3n-2)}{6} C^2_2(SU(2)_I)_{\bold{2}} + \frac{n(4n^2-6n+3)}{6} C^2_2(SU(2)_F)_{\bold{2}} \\ \nonumber & + & \frac{(10n+1)( 7 p^2_1 (T) - 4 p_2(T))}{1920} .
\eea

Here we use $SU(2)_I$ for the $SU(2)$ global symmetry of $SU(2) \times E_7$. For $n=2$, corresponding to $SO(7) \times E_7$ conformal matter, these can be combined to give:

\bea
I_{SO(7) \times E_7} & = & \frac{119}{24} C^2_2(R) - \frac{49}{48} p_1 (T) C_2(R) + \frac{1}{6} p_1 (T) C_2(SO(7))_{\bold{8}} + \frac{1}{24} p_1 (T) C_2(E_7)_{\bold{56}} \\ \nonumber & - & \frac{1}{3} C_2(R) C_2(E_7)_{\bold{56}} - \frac{3}{2} C_2(R) C_2(SO(7))_{\bold{8}} + \frac{1}{144} C^2_2(E_7)_{\bold{56}} + \frac{1}{24} C_2(E_7)_{\bold{56}} C_2(SO(7))_{\bold{8}} \\ \nonumber & + &  \frac{5}{24} C^2_2(SO(7))_{\bold{8}} - \frac{1}{6} C_4(SO(7))_{\bold{8}} + \frac{7( 7 p^2_1 (T) - 4 p_2(T))}{640} .
\eea

\subsubsection{Computing $4d$ anomalies from $6d$}

We can use the anomaly polynomial to compute the anomalies of $4d$ theories resulting from the compactification of the $6d$ SCFT on a Riemann surface. Here we shall take the Riemann surface to be a torus. We shall also consider non-zero flux in $SU(2)_E$, the one associated with the singularity. Specifically, we take $C_2(SU(2)_E)_{\bold{2}} = - C^2_1 (U(1)_E)$, $C_1 (U(1)_E) = - z t + \epsilon U(1)^{6d}_R + C_1 (U(1)^{4d}_E)$ so that $\int_{\Sigma} C_1 (U(1)_E) = -z$.

Performing the calculation we find, for the $SU(2) \times E_7$ SCFT:

\be
a = \frac{(18n-5)\sqrt{18n-5}|z|}{24} , \quad c = \frac{(18n-1)\sqrt{18n-5}|z|}{24}, \label{acEseven}
\ee

while for $SO(16)$ the result is:

\be
a = \frac{(9n+2)\sqrt{9n+2}|z|}{6 \sqrt{2}} , \quad c = \frac{(9n+4)\sqrt{9n+2}|z|}{6 \sqrt{2}} . \label{acSO}
\ee

 Finally we consider the special cases in a bit more detail. First the $n=1$ case for the $SO(16)$ SCFT gives the $(D_5, D_5)$ conformal matter. In that case the symmetry is enhanced to $SO(20)$, the flux to $SU(2)_E$ describes a breaking of $SO(20)$ to $U(1) \times SU(2) \times SO(16)$, and our results here match those of \cite{KRVZ2}. 

The $n=2$ case for the $SU(2) \times E_7$ SCFT gives the $SO(7) \times E_7$ conformal matter. Here the three $SU(2)$ symmetries enhance to $SO(7)$, where flux to $SU(2)_E$ describes the breaking of $SO(7)\rightarrow U(1)\times SU(2)\times SU(2)$. From (\ref{acEseven}) we find that:

\be
a = \frac{31\sqrt{31}|z|}{24} , \quad c = \frac{35\sqrt{31}|z|}{24}, \label{acSOEs}
\ee

We can also consider more general fluxes inside $SO(7)$. For example, we can consider flux breaking $SO(7)\rightarrow U(1)\times USp(4)$, for which we find:

\be
a = \frac{31\sqrt{31}|z|}{12\sqrt{7}} , \quad c = \frac{35\sqrt{31}|z|}{12\sqrt{7}}, \label{acSOEs}
\ee

To study more general cases, we shall introduce a basis of three fluxes labeled $(a, b, c)$. We shall use essentially the same basis used in \cite{RVZ}. This basis is defined so that  $a$ and $b$ are associated with fluxes in the Cartans of $SU(2)_E$ and $SU(2)_I$ respectively while $c$ is associated with flux in the Cartan of $SU(2)_F$. We also normalize the Cartans so that the charges in the vector representation are unity. So using fugacities $e$, $i$, $f$ for the associated $U(1)$'s:

\be
\bold{7} \rightarrow 1 + f + \frac{1}{f} + (e + \frac{1}{e})(i + \frac{1}{i}) . 
\ee 

Note that, with this normalization, the fluxes are quantized in the following manner. The charges appearing for $SU(2)_E$ and $SU(2)_I$ in both the vector and the spinor are both $1$ so naively $a, b$ must be integers. For $SU(2)_F$, however, while the charge appearing in the vector is $1$, the spinor has states with charge $\frac{1}{2}$, so naively $c$ must be an even integer. Yet, the vector is consistent with $a, b$ being both half-integer while the spinor is not, but the spinor will be consistent if $c$ is an odd integer. Thus, we conclude that the flux is quantized so that either $c$ is an even integer and $a, b$ integers, or $c$ is an odd integer and $a, b$ are both half-integer. For generic fluxes, where the global symmetry is broken to $U(1)^3$, these are the only possible values. Nevertheless, if some non-abelian symmetry is present, it is possible to support other values of flux, which are made consistent with central fluxes in the non-abelian group, see appendix C in \cite{KRVZ} for a general discussion. We shall also see examples of this later in various $4d$ tubes.

We can next consider compactification with general flux $(a, b, c)$. We shall summarize the anomalies in terms of the traces for $R = U(1)^{6d}_R + \epsilon_e SU(2)_E + \epsilon_i SU(2)_I + \epsilon_f SU(2)_F$. We then find:

\bea
 Tr(R^3) & = & \frac{7}{2} c \epsilon^3_f + 9 (a \epsilon_e + b \epsilon_i)(\epsilon^2_f-4) + 9 c \epsilon_f (\epsilon^2_i+\epsilon^2_e-2) + 16 (a \epsilon^3_e + b \epsilon^3_i) + 12 \epsilon_i \epsilon_e (b \epsilon_e + a \epsilon_i) , \nonumber \\  
 Tr(R) & = & 8 (c \epsilon_f + 2 b \epsilon_i + 2 a \epsilon_e ) .
\eea

\subsubsection{Contribution of the punctures to the anomalies}

So far we assumed that the surface is a torus, and we now wish to relax that assumptions and also consider the case of tubes. The important difference between the two is that the latter contains punctures. These have degress of freedom living on them whose contribution to the anomalies is needed to also be taken into account. We have previously mentioned that the punctures can be described by boundary conditions on $5d$ gauge theories, and these can be used to compute their contribution to the anomalies as we shall now detail. This method was originally used in \cite{KRVZ} and later also in \cite{KRVZ1,KRVZ2} for the determination of the puncture contribution to the anomalies.

There are two notable sources for the $4d$ anomaly contribution of the punctures. One, is the contribution from the fermions in multiplets given a Nuemann boundary conditions. The second is from Chern-Simons terms that may exist in the $5d$ gauge theory. Here we shall not need the latter contribution and so concentrate only on the former.

The general idea is that each multiplet given a Nuemann boundary conditions contribute half the expected anomaly. This is motivated by the following 'thought experiment'. Consider reducing a free $5d$ fermion on an interval. At the $4d$ theory on the boundary, we can split the $5d$ fermion to two $4d$ Weyl fermions of opposite chirality. We can now give Dirichlet boundary conditions for one chirality and Nuemann for the other at the two sides of the interval. Reducing to $4d$ we expect to be left only with the $4d$ Weyl fermion that received Nuemann boundary conditions. The $4d$ anomaly of this fermion should then match the anomaly inflow contribution from both punctures, and, therefore, each should contribute half of the anomaly.

Here we shall mostly be concerned about the case where the $5d$ gauge theory description is $SU(N)_0+2AS+8F$. For this case we need to consider the contribution of the fields receiving Nuemann boundary conditions in the vector and hypermultiplets. For the vector multiplet, we are giving Nuemann boundary conditions to the adjoint chiral. This field is only charged under the $SU(N)$ gauge symmetry, in the adjoint representation, and under the $SU(2)_R$. The former is the puncture symmetry in $4d$, while the latter is broken by the flux to $U(1)^{6d}_R$. As the fermions in the chiral adjoint and the vector multiplet form an $SU(2)_R$ doublet, and as the latter always as R-chrage $1$ under the $\mathcal{N}=1$ $U(1)$ R-symmetry, we conclude that the fermion in the chiral adjoint has charge $-1$ under $U(1)^{6d}_R$. Thus, we find it contribute to the anomalies:

\be
Tr(U(1)^{6d}_R) = Tr((U(1)^{6d}_R)^3) = - \frac{(N^2-1)}{2}, \quad Tr(U(1)^{6d}_R SU(N)^2) = - \frac{N}{2}.
\ee 

 We next want to consider the contribution of the hypermultiplets. Here, we are giving a Nuemann boundary conditions to one of the chiral fields in the hypermultiplets. The fermion in that chiral field is then charged under the $6d$ global symmetry and the $SU(N)$ gauge symmetry in some representation $\bold{R}$. For simplicity, we consider the case of only one of these, charged under some $U(1)_F$ with charge $q$, where the generalization to more general cases being immediate. We then expect the following contribution to the anomalies:

\bea
& & Tr(U(1)_F) = \frac{q}{2}, \quad Tr(U(1)^3_F) = \frac{q^3}{2}, \\ \nonumber & & Tr(U(1)_F SU(N)^2) = \frac{q T_{\bold{R}}}{2}, \quad Tr(SU(N)^3) = \frac{D_{\bold{R}}}{2},
\eea      
where we use $T_{\bold{R}}$ and $D_{\bold{R}}$ for the second and third Dynkin indices of the representation $\bold{R}$, respectively.

Combining both terms gives the puncture contribution to the anomalies.

\section{$4d$ analysis}

\begin{figure}
\center
\includegraphics[width=0.9\textwidth]{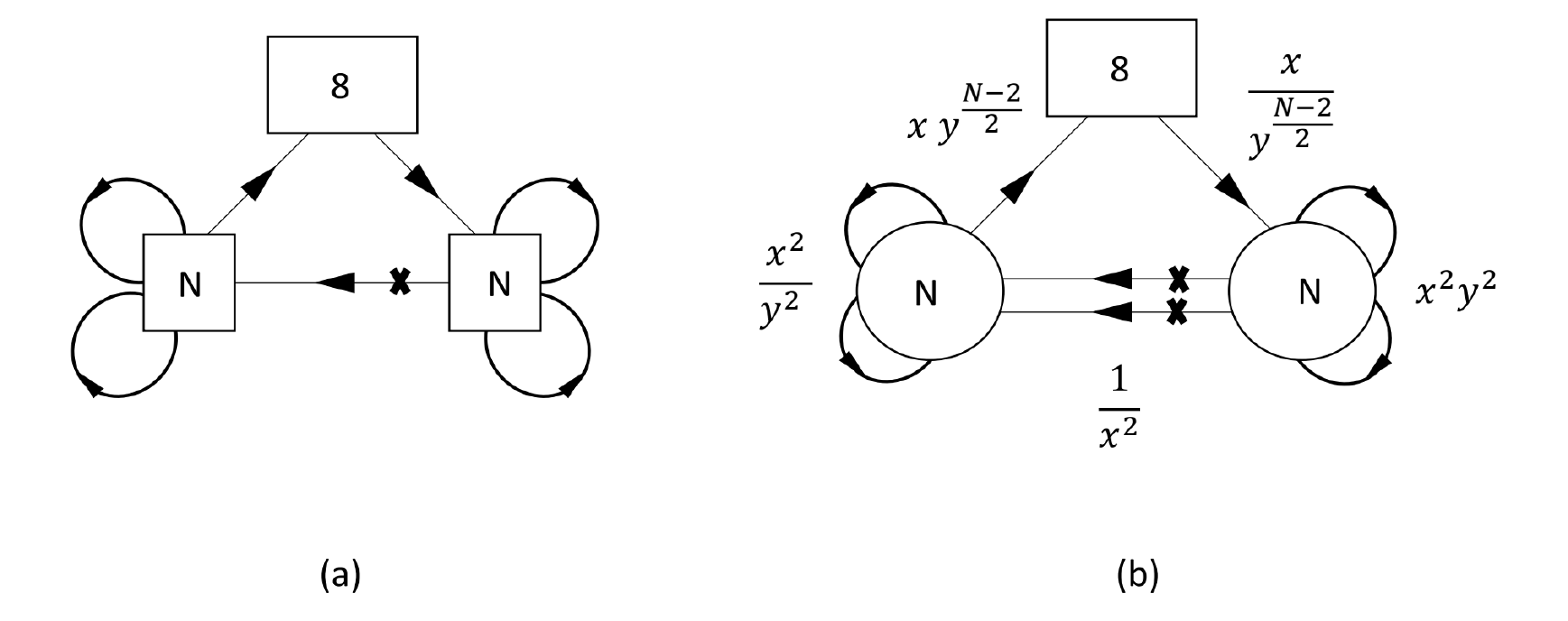} 
\caption{$4d$ theories associated with the compactification of the $SU(2) \times E_7$ SCFT for $N=2n$ and of the $SO(16)$ SCFT when $N=2n+1$. In (a) the compactification surface is a sphere with two maximal punctures, and the flux is $z=\frac{1}{2}$ in $SU(2)_E$. In (b) the compactification surface is a torus, and the flux is $z=1$ in $SU(2)_E$. The arrows from a group to itself are all antisymmetric or conjugate antisymmetric chiral fields, where the representation is such that they can form a gauge invariant with an antisymmetric combination of the bifundamentals. In both theories there is a cubic superpotential along the triangle and also a quartic one connecting the antisymmetrics on the two sides via two bifundamentals.}
\label{TubeaThr}
\end{figure}

In this section we shall present the $4d$ theories we associate with the $6d$ compactifications we presented. These are conjectured using the $5d$ domain wall approach where the domain wall is taken to extrapolate between two $SU(N)$ $5d$ descriptions. Consider the theory shown in figure \ref{TubeaThr} (a). We associate this theory with a compactification on a sphere with two maximal punctures, with flux $z=\frac{1}{2}$ in $SU(2)_E$, of the $SU(2) \times E_7$ SCFT for $N=2n$ and of the $SO(16)$ SCFT when $N=2n+1$\footnote{The choice of flux for this tube was determined, as will be supported momentarily, by matching symmetries and anomalies. In principle, it should be possible to determine it by understanding the variable holonomy, like in \cite{KRVZ}. We won't consider this here though.}. Here the two maximal punctures have opposite colors.

One can see that this theory is essentially made from free fields with the $5d$ matter content, not killed by the boundary conditions, on both sides\footnote{Note that the adjoint chiral in the $\mathcal{N}=2$ vector multiplet does not appear in the tube, despite being given a Nuemann boundary conditions. This is in fact quite common in these constructions, see \cite{KRVZ,KRVZ1,KRVZ2}. In cases where the domain walls are well understand, like in \cite{KRVZ1}, then this is attributed to it being given Dirichlet boundary conditions on the domain wall.} connected by a bifundamental, with a singlet field flipping its baryon. The latter fields are attributed to the domain wall. Next we shall support this conjectural tube by a variety of tests. 

 We begin by combining two tubes to form a torus, using the gluing rules explained in the previous section. This leads to the theory in figure \ref{TubeaThr} (b). As we have glued two tubes with flux $z=\frac{1}{2}$ to form a torus, we expect this theory to be associated with a torus compactification with with flux $z=1$ in $SU(2)_E$. We first note that the gauge anomalies vanish where the contributions of the two antisymmetrics and eight flavors canceling against that of the bifundamentals, which is one non-trivial consistency check. We have two anomaly free $U(1)$'s which we denote as $U(1)_x$ and $U(1)_y$. Additionally we have the $SU(8)$ rotating the $8$ flavors and the $SU(2)$ rotating the $2$ antisymmetrics. Overall, the theory has visible global symmetry $U(1)_x\times U(1)_y\times SU(2) \times SU(8)$. Here $U(1)_x$ is related to the Cartan of $SU(2)_E$, and the $SU(2)$ is $SU(2)_F$. When $N$ is even then $SU(8)$ is the maximal subgroup of $E_7$ and $U(1)_y$ the Cartan of $SU(2)_I$ while when $N$ is odd then $U(1)_y \times SU(8)$ is the maximal subgroup of $SO(16)$. There is also a natural R-symmetry, under which the bifundamentals have R-charge $0$, the flippers R-charge $2$ and everything else R-charge $1$, which is identified with $U(1)^{6d}_R$. These identifications are supported by anomaly matching between $6d$ and $4d$. 

Both $SU(N)$ groups are asymptotically free so the theory can exhibit interesting dynamics in the IR. Using a maximization we can determine the superconformal R-symmetry. The only $U(1)$ that can mix is $U(1)_x$ so we take $U(1)^{sc}_R = U(1)^{6d}_R + \alpha U(1)_x$. Performing the a maximization we find $\alpha=-\frac{\sqrt{9N-5}}{6N}$. With this R-symmetry, the a and c central charges match (\ref{acEseven}) and (\ref{acSO}). We can also compare the other anomalies and find they match according to the mapping we prescribed. More preciously, we need to take $\bold{2}_{SU(2)_E} = x^N + \frac{1}{x^N}$, $\bold{2}_{SU(2)_I} = y^N + \frac{1}{y^N}$ for the $N$ even and $\bold{16}_{SO(16)} = y^{\frac{N}{2}} \bold{8}_{SU(8)} + \frac{1}{y^{\frac{N}{2}}}\bar{\bold{8}}_{SU(8)}$ for $N$ odd, up to charge conjugation of $SU(8)$.

We next consider some aspects of the spectrum of the theories. Having found the superonformal R-smmetry, we 
first check whether there are operators below the unitary bound. We find that only the singlets go below the unitary bound. Thus, it is reasonable that these decouple and become free fields. Preforming the a maximization without them we now find $\alpha = -\frac{(3N+1)}{12N}$, using which we observe that all gauge invariant operators are above the unitary bound. Therefore, it is reasonable that the $4d$ theory in figure \ref{TubeaThr} (b) go to a $4d$ SCFT and two decoupled free chiral fields.

Finally, we look at some of the gauge invariant states in the theory in order to try and match them to the $6d$ spectrum. First we have the two singlets, which are identified as coming from the negatively charged generator of $SU(2)_E$. We have two such operators in accordance with the logic of \cite{BRZ} (see also appendix E in \cite{KRVZ}). The basic other gauge invariants are baryons made from the flavors and the antisymmetrics. Out of these, the ones with lowest R-charge are those made of the largest number of antisymmetrics. In the $N$ even case, corresponding to the compactification of the $SU(2)\times E_7$ SCFT, we first have the one made just from $\frac{N}{2}$ antisymmetrics. This gives a gauge invariant that contribute to the index the term: $x^N (y^N + \frac{1}{y^N}) \chi[\frac{N}{2} + 1]_{SU(2)} (p q)^{\frac{N}{4}}$, where here we use the $6d$ R-symmetry. Next we have the one made from two flavors and $\frac{N}{2}-1$ antisymmetrics. This gives a gauge invariant that contribute to the index the term: $x^N (\chi[\bold{28}]_{SU(8)} + \chi[\bar{\bold{28}}]_{SU(8)}) \chi[\frac{N}{2}]_{SU(2)} (p q)^{\frac{N+2}{4}}$. These match the $SU(2)_E$ charged states expected from the $6d$ SCFT. We also expect ones with the same charges save $U(1)_x$ which should be $-N$.  

In the $N$ odd case, now corresponding to the compactification of the $SO(16)$ SCFT, we now must use an odd number of flavors. The simplest one is made from just one flavor and $\frac{N-1}{2}$ antisymmetrics. This gives a gauge invariant that contribute to the index the term: $x^N (y^{\frac{N}{2}}\chi[\bar{\bold{8}}]_{SU(8)} + \frac{1}{y^{\frac{N}{2}}}\chi[\bold{8}]_{SU(8)}) \chi[\frac{N+1}{2}]_{SU(2)} (p q)^{\frac{N+1}{4}}$. This correctly matches the properties expected from the sole $SU(2)_E$ charged state expected from the $6d$ SCFT. We again expect also ones with the same charges but with $U(1)_x$ charge of $-N$.

This observation is not as accurate as the supercoformal index so in particular we have not determined the number of such operators and so also have not ruled out the possibility of cancellations. Still we observe operators which can be naturally matched with those expected from $6d$, which in turn is a strong indication that this theory has a $6d$ origin.  

We now return to the tube. After having determined the mapping of the symmetries between $4d$ and $6d$, we can also compare the anomalies of the tube against the $6d$ expectation. Note that the tube has an additional $U(1)$ that is accidental from the $6d$ viewpoint. This makes it easier to compare anomalies for closed surfaces first. The charges of the fields in the tube, under the $6d$ symmetries, can be easily mapped from those in figure \ref{TubeaThr} (b), and with the charge map determined, it is straightforward to compare the anomalies using the results of the previous section. We indeed find that all anomalies match. For example, consider the $Tr(SU(N)^3)$ anomaly associated with the left puncture. As the symmetry is associated to a puncture, it receives contribution only from the punctures. We see that:

\be
Tr(SU(N)^3)_{6d} = -4 - N + 4 = -N ,
\ee

\be
Tr(SU(N)^3)_{4d} = -8 - 2 N + 8 + N = -N ,
\ee
and these indeed match.

We can also consider gluing multiple tubes to form torus with higher values of flux. By virtue of the matching of the tube anomalies, the anomalies of the resulting $4d$ theories are guaranteed to match those expected from $6d$. One interesting aspect is what happens when we glue an odd number of tubes, the simplest case being gluing a tube to itself.

Consider closing the tube in figure \ref{TubeaThr} (a) on itself to form a torus compactification with flux $z=\frac{1}{2}$. If we try to do this according to the prescription set up previously, that is we gauge both $SU(N)$ global symmetries by an $\mathcal{N}=1$ $SU(N)$ vector multiplet, and try to add appropriate chiral fields, we encounter a problem. Basically, because the punctures have different colors we naively need to introduce chiral fields with different $SU(N)$ representations, fundamental versus antifundamental, depending on which of the puncture we consider.

The solution we find for this problem is that here the gluing must be done with the two $SU(N)$ groups identified with a charge conjugation. When glued in this manner the additional chiral fields can be added and coupled to the matter via a superpotential consistently. We next consider the resulting theory and show that the results are consistent with it being a $6d$ torus compactifications with flux $z=\frac{1}{2}$.   

\begin{figure}
\center
\includegraphics[width=0.3\textwidth]{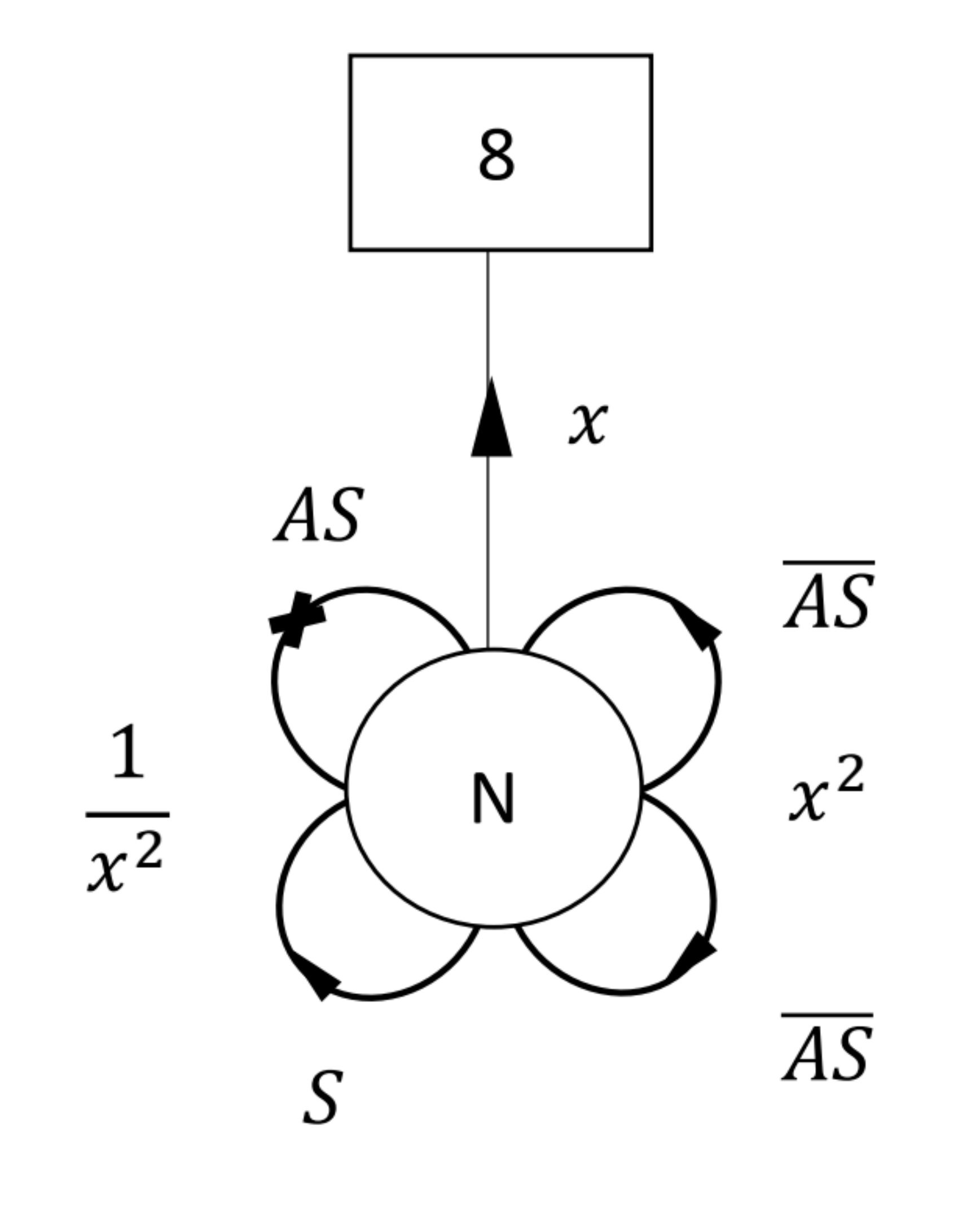} 
\caption{The $4d$ theories associated with the torus compactification of the $SU(2) \times E_7$ SCFT for $N=2n$ and of the $SO(16)$ SCFT when $N=2n+1$ with flux $z=\frac{1}{2}$ in $SU(2)_E$. Here the $8$ flavors are in the antifundamental of the $SU(N)$ gauge symmetry. The lines from the group to itself stands for symmetric, denoted by $S$, antisymmetric, denoted by $AS$, or their conjugate representation, denoted by a bar. The square with $8$ represents an $SO(8)$ or $USp(8)$ global symmetry, depending on the exact choice of superpotential.}
\label{TubeGlued}
\end{figure}

The resulting theory is shown in figure \ref{TubeGlued}. The two conjugate antisymmetric chirals on the right come from the analogue chirals in the tube and are rotated by the $4d$ version of $SU(2)_F$. The antisymmetric and symmetric chirals on the left come from the bifundamental in the tube. There is also a chiral singlet, that flipped the bifundamental baryon in the tube, that now flip baryons made from $N$ symmetric or antisymmetric chirals. In the figure it is denoted as an $X$ on the symmetric chiral, although we stress that it is coupled also to analogous combinations made from the antisymmetric. There is also a quartic superpotential coupling the two conjugate antisymmetrics to the symmetric or antisymmetric chirals and a cubic superpotential coupling two antifundamental chirals to the symmetric or antisymmetric chirals, on which we shall elaborate more next.

When gluing the tube to itself we were forced to break $U(1)_y$ and the $SU(8)$ rotating the flavor though the supperpotential. The latter specifically, is broken by the cubic superpotential coupling two flavors with the symmetric or antisymmetric hyper. This superpotential comes from the one along the triangle in the tube. The one coupling through the symmetric breaks $SU(8)$ to $SO(8)$ while the one coupling through the antisymmetric breaks $SU(8)$ to $USp(8)$. Both superpotential are consistent with the $6d$ symmetry, and if the theory flows to an SCFT, should both be marginal. In that case we can turn then both on, which breaks the flavor symmetry to $U(1)^4$. Thus, we expect a conformal manifold on a generic point of which the symmetry is broken to $U(1)^4$, but with special lines along which the symmetry is enhanced to $SO(8)$, $USp(8)$ and various combinations of $SO$ and $USp$ when we couple part of the flavors using the symmetric and the other using the antisymmetric. Of coarse it is possible that some of these marginal operators are actually marginally irrelevant. 

Besides what remains of the $SU(8)$, the theory also has $U(1)_x$ as a global symmetry, with the charges shown in the figure using fugacities, and a $U(1)_R$ symmetry, which we identify with $U(1)^{6d}_R$, under which the flip field has R-charge $2$, the symmetric and antisymmetric chirals have R-charge $0$, and the rest R-charge $1$.

As a first test, we can compare the anomalies of this $4d$ theory against those expected from $6d$, using the previous mapping. Indeed, we find that all the anomalies match. As an example, consider the conformal anomalies. Taking a trial R-symmetry as $U(1)^{6d}_R + \alpha U(1)_x$ and using a maximization we find $\alpha = - \frac{\sqrt{9N-5}}{6N}$. We can now use this to evaluate the central charges finding:

\be
a = \frac{(9N-5)\sqrt{9N-5}}{48}, \quad c = \frac{(9N-1)\sqrt{9N-5}}{48} ,
\ee     
which matches (\ref{acEseven}), (\ref{acSO}) for $z=\frac{1}{2}$. Like in the previous case, the chiral singlet drops below the unitarity bound and so these values of the central charges are not the actual central charges of the IR theory, but as the $6d$ theory is insensitive to these issues, they still work well for anomaly matching between $6d$ and $4d$. Again we can repeat the analysis, but now decoupling the singlet, finding $\alpha=-\frac{(3N+1)}{12N}$. Now, we find that all gauge invariant operators are above the unitarity bound, and so it is plausible that this theory flows to an SCFT plus a decoupled chiral field.

Finally we need to discuss the global symmetry and fractional flux. The fractional flux here is only possible as the global symmetry of the SCFTs appear to be not simply connected, as pointed out in the previous section. In this case, the phase from the fractional flux can be canceled by a pair of almost commuting holonomies, see section 5 in \cite{BHMRTZ} and appendix C in \cite{KRVZ}. This is expected to break part of the global symmetry.

Let us first consider the case of the $SO(16)$ SCFT. Here the flux must be accompanied by a pair of holonomies commuting in $Spin(16)/Z_2$ but not in $Spin(16)$, where the $Z_2$ element is the center that acts non-trivialy on the vector and one of the spinors. This is the so called compactifications without vector structure studied in \cite{Wit}. As pointed out in the reference, there is in general a choice of such holonomies having the following properties. A generic choice breaks $SO(16)$ to $U(1)^4$, but there are special choices where the symmetry enhances to $SO(8)$, $USp(8)$ or various $USp$, $SO$ combinations. The holonomies corresponding to these choices can be continuously transformed to one another, in other words, there are no disconnected components.

Here the choice of holonomies of the flavor symmetry on the torus is expected to map to marginal operators in the resulting $4d$ theory, see \cite{BTW}. Therefore, we see that the resulting structure agrees with what we observe in $4d$, at least qualitatively. The almost commuting holonomies break $SO(16)$ to at least $U(1)^4$. The holonomy space is mapped to the conformal manifold, where at different points the $U(1)^4$ is enhanced to various symmetries, specifically $SO(8)$, $USp(8)$ or various $USp$, $SO$ combinations. It should be noted that we have not compared the dimension of the conformal manifold against the $6d$ expectation based on holonomies, or even showed that the the operators are exactly marginal, hence this being a qualitative matching. We should also mention that the relation between the holonomy space and the conformal manifold is known to not always hold for tori with low values of flux, like the theory we consider here, see \cite{RVZ,BHMRTZ,KRVZ}, so that even if there happens to be a disagreement, it is while disappointing, not entirely unexpected.    

Finally, we consider the case of the $SU(2)\times E_7$ SCFT. The same story also holds here, but now for consistency with the fractional flux, we need two almost commuting holonomies both in $SU(2)_I$ and the $E_7$. The holonomies in $SU(2)_I$ break it completely, which agrees with the breaking of $U(1)_y$ in the $4d$ field theory. The holonomies in $E_7$ break it generically to $U(1)^4$, which can be enhanced for special choices to among others $SO(8)$ and $USp(8)$. Interestingly, the largest group one can preserve is $F_4$, see appendix C in \cite{KRVZ}. Thus, there could be a line on the conformal manifold with $F_4$ global symmetry, where the doubt here is due to the cautionary remarks in the end of the previous paragraph.

\subsection{Special cases} 

There are two special values of $N$, where the symmetry enhances in $6d$ and some special features appear in the $4d$ theory. One case is the $N=3$ case where the $6d$ SCFT is the $(D_5,D_5)$ conformal matter studied in \cite{KRVZ1}. The tube we find indeed matches one of the tubes found there. The second special case is $N=4$, which we now discuss.

\subsubsection{$SO(7) \times E_7$ conformal matter}

The special feature of the $N=4$ case is that the $SU(2)\times SU(2)\times SU(2)$ symmetry is enhanced to $SO(7)$. This leads us to suspect that we should be able to find tubes associated with other choices of flux inside the $SO(7)$ global symmetry group. The reason for this is the following. When the symmetry is a large group, then we should be able to rotate the flux inside the larger group using various Weyl transformations. While these give the same flux, only embedded differently in the larger group, we can consider gluing together two domain walls associated with this same flux but with different embeddings. The resulting domain wall is then associated with a different flux.

Indeed, in the previous studies of compactifications of $(1,0)$ SCFTs on torus with fluxes this as been used to generate additional domain walls with more general fluxes\cite{KRVZ,KRVZ1,KRVZ2}. The structure that was observed there was that these generalized domain walls have matter content similar to just the gluing of the basic domain wall to itself, but with some of the matter chiral fields projected out. We next use this observation to conjecture, and then test, these additional domain walls.  

For generic $N$ such a program appears unfeasible, the reason being that the matter content is extremely restricted by gauge anomaly cancellation. However, for $N=4$, we have $SU(4)$ groups for which the antisymmetric representation is real. This in particular means that they do not contribute to the gauge anomalies so theories with some of them removed still make sense. This is in accordance with our previous observation, and leads us to suspect that gluing two of the tubes in figure \ref{TubeaThr} (a), but with some of the antisymmetrics removed, may lead to more general tubes.

\begin{figure}
\center
\includegraphics[width=1\textwidth]{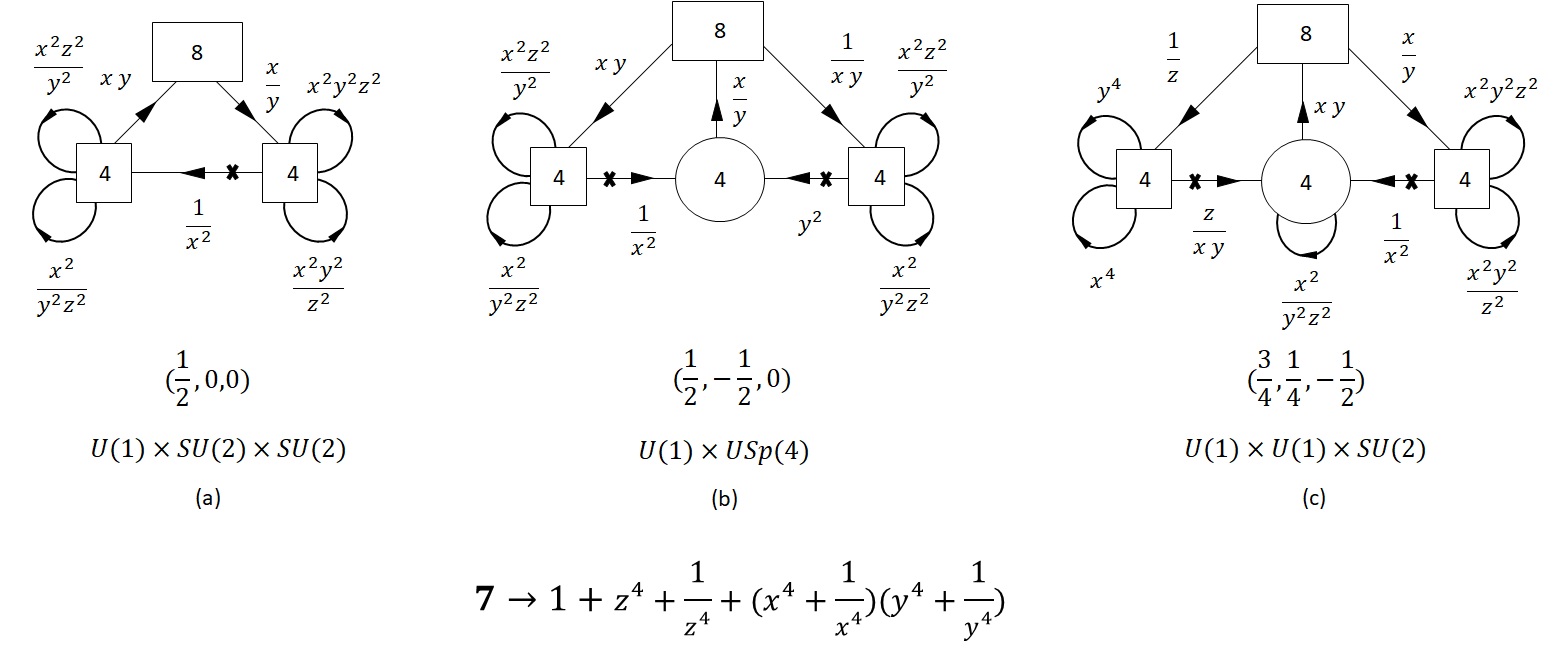} 
\caption{$4d$ tubes for the $SO(7) \times E_7$ conformal matter. Below each tube is the flux associated with it and the global symmetry it preserves (using the notation of \cite{RVZ}). Finally we show how the $SO(7)$ is built in term of the symmetries we defined in $4d$.}
\label{TubesESOCM}
\end{figure}

Analyzing all cases we get the three tubes in figure \ref{TubesESOCM}. The one in (a) is just the one presented previously specialized to the case of $N=4$. The ones in (b) and (c) are given by connecting two tubes but removing some of the antisymmetrics. By analyzing anomalies of the theories we get by connecting two tubes, as well as the tubes themselves, we get the mapping of fluxes. These span the complete $SO(7)$ flux basis.   

We can test this by connecting two tubes to form a closed surface. In this case we can compute the anomalies and compare against the $6d$ results. We can also compute the anomalies of the tubes themselves and compare. Furthermore, we can consider combining two tubes of type (a) with two tubes of type (b) or (c) and check the anmalies of the resulting theories. In all cases we checked we have found agreement between the $6d$ and $4d$ results. 

Finally we can consider the superconfomal index. Since the $N=4$ case is the simplest one not previously considered, it is a good case for these type of computations. As an example we take the theory in figure \ref{TubeaThr} (b) for $N=4$, which is the same as the one we get by gluing together two tubes of the type shown in figure \ref{TubesESOCM} (a). We shall take $U(1)^{6d}_R - \frac{1}{5} U(1)_x$ as the R-symmetry. Since $\frac{\sqrt{31}}{24}-\frac{1}{5} \approx 0.032$, it is close to the actual superconformal R-symmetry. We shall also ignore the singlets as these hit the unitary bound and so should decouple in the IR. Then for the index we find:

\bea
1 & + & x^4 (p q)^{\frac{3}{5}} (y^4 + \frac{1}{y^4})(1 + z^4 + \frac{1}{z^4}) + 5 \frac{(p q)^{\frac{4}{5}}}{x^8} + p q (z^4 + \frac{1}{z^4}) \\ \nonumber & + & x^4 (p q)^{\frac{3}{5}} (y^4 + \frac{1}{y^4})(1 + z^4 + \frac{1}{z^4})(p + q) + x^4 (p q)^{\frac{11}{10}} (z^2+\frac{1}{z^2})(\chi[\bold{28}]_{SU(8)} + \chi[\bar{\bold{28}}]_{SU(8)}) \\ \nonumber & + & x^8 (p q)^{\frac{6}{5}} \left( (y^8 + 1 + \frac{1}{y^8})(2 + z^4 + \frac{1}{z^4}+ z^8 + \frac{1}{z^8}) + 1 + z^4 + \frac{1}{z^4} \right) + ...
\eea

First we note that the index indeed form characters of the expected $6d$ symmetry. Particularly, $U(1)_y$ seem to form an $SU(2)$ with $\bold{2}_{SU(2)_y} = y^4 + \frac{1}{y^4}$ and $SU(8)$ seem to enhance to $E_7$ with $\bold{56}_{E_7} = \chi[\bold{28}]_{SU(8)} + \chi[\bar{\bold{28}}]_{SU(8)}$.

Second, we can identify some of the states here as contributions of fundamental $6d$ states. Recall that under the decomposition of $SO(7)\rightarrow U(1)\times SU(2)\times SU(2)$, we have $\bold{21} \rightarrow (\bold{3},\bold{1})^0 + (\bold{1},\bold{3})^0 + (\bold{1},\bold{1})^0 + (\bold{1},\bold{1})^{\pm 2} + (\bold{3},\bold{2})^{\pm 1}$. Then the first term in the index matches the contribution expected from the last term in the decomposition (the second to last term contribution to the relevant operators is matched by the singlets). We also have $\bold{8} \rightarrow (\bold{2},\bold{2})^0 + (\bold{2},\bold{1})^{\pm 1}$, under this decomposition. We can then identify the state at order $(p q)^{\frac{11}{10}}$ with the contribution of the $(\bold{8},\bold{56})$ state of the $6d$ SCFT. This confirms the crude analysis, that we previously preformed generically, for this special case.


We can also consider the superconformal index of the theory when we glue the tube to itself, that is the one in figure \ref{TubeGlued} for $N=4$. We can perform the evaluation with the same non-superconformal R-symmetry, $U(1)^{6d}_R - \frac{1}{5} U(1)_x$. Here we shall again ignore the singlet, as it decouples in the IR. Furthermore, we shall take the symmetry rotating the $8$ flavors to be $SO(8)$. Evaluating the index we find it to be:

\bea
1 & + &  \frac{(p q)^{\frac{2}{5}}}{x^4} +(p q)^{\frac{1}{2}} (z^2 + \frac{1}{z^2}) + x^4 (p q)^{\frac{3}{5}} (1 + z^4 + \frac{1}{z^4}) + 3 \frac{(p q)^{\frac{4}{5}}}{x^8} + \frac{(p q)^{\frac{9}{10}}}{x^4} (z^2 + \frac{1}{z^2}) \\ \nonumber & + & p q (2 z^4 + 1 + \frac{2}{z^4}) + x^4 (p q)^{\frac{3}{5}}(1 + z^4 + \frac{1}{z^4})(p + q) + x^4 (p q)^{\frac{11}{10}} (z^2+\frac{1}{z^2})(z^4+\frac{1}{z^4} + \chi[\bold{28}]_{SO(8)}) \\ \nonumber & + & (p q)^{\frac{6}{5}} \left( \frac{3}{x^{12}} + ( 2 + z^4 + \frac{1}{z^4} + z^8 + \frac{1}{z^8})x^{8} \right) + ...
\eea 

The interesting thing here is the $(p q)^{\frac{11}{10}}$ terms that are in $SO(8)$ characters. As previously mentioned, we can break the $SO(8)$ group down to its maximal torus, by changing the superpotential, which is mapped to the exact choice of almost commuting pairs. From the latter viewpoint we expect a point with $F_4$ global symmetry, which we do not see in the Lagrangian.

Now, consider choosing the superpotential such that two quarks are paired with the symmetric and the remaining six with the antisymmetric. This should lead to a symmetry of $SO(2)\times USp(6)$, under which the states in the $\chi[\bold{28}]_{SO(8)}$ become instead $2\bold{1}^0 + \bold{14}^0 + \bold{6}^1 + \bold{6}^{-1}$. Curiously this can be cast into the $\bold{26}$ of $F_4$ where here the $SO(2)\times USp(6)$ is embedded in the $SU(2)\times USp(6)$ maximal subgroup of $F_4$. This is consistent with our $6d$ expectations. 

\section{Mass deformed theories}

In this section we discuss $4d$ theories related to compactifications of $6d$ theories not through a direct reduction, but rather one also involving a mass deformation. The way we approach these theories will be similar to the steps used in motivating the previous theories, particularly relying on the $5d$ gauge theory description. As previously mentioned, the $6d$ SCFTs considered here all have a $5d$ description as an $SU$ type gauge theory with hypermultiplets in the fundamental and antisymmetric representations. The previous theories were then motivated using domain walls between two such $5d$ descriptions. However, the $6d$ SCFTs also have other descriptions, in particular, they possess one involving two $USp$ gauge groups. We can then consider a $5d$ domain wall extrapolating between the two descriptions, that is haing the $5d$ $SU$ description on one side, and the $USp\times USp$ quiver description on the other. When used in torus compactifications this is expected to give $4d$ theories involving two $USp$ groups on one side and an $SU$ group with antisymmetrics on the other. As we shall see in this section, this leads to rather interesting theories that, while not related to direct compactification of the relevant $6d$ SCFTs, are connected to them by a mass deformation.

The behavior of the $4d$ theory is quite different depending on whether we consider the $SO(16)$ or the $SU(2)\times E_7$ SCFT, and thus we shall discuss each case separately, beginning with the $SO(16)$ case.   

\subsection{$SO(16)$ SCFT}

\begin{figure}
\center
\includegraphics[width=0.42\textwidth]{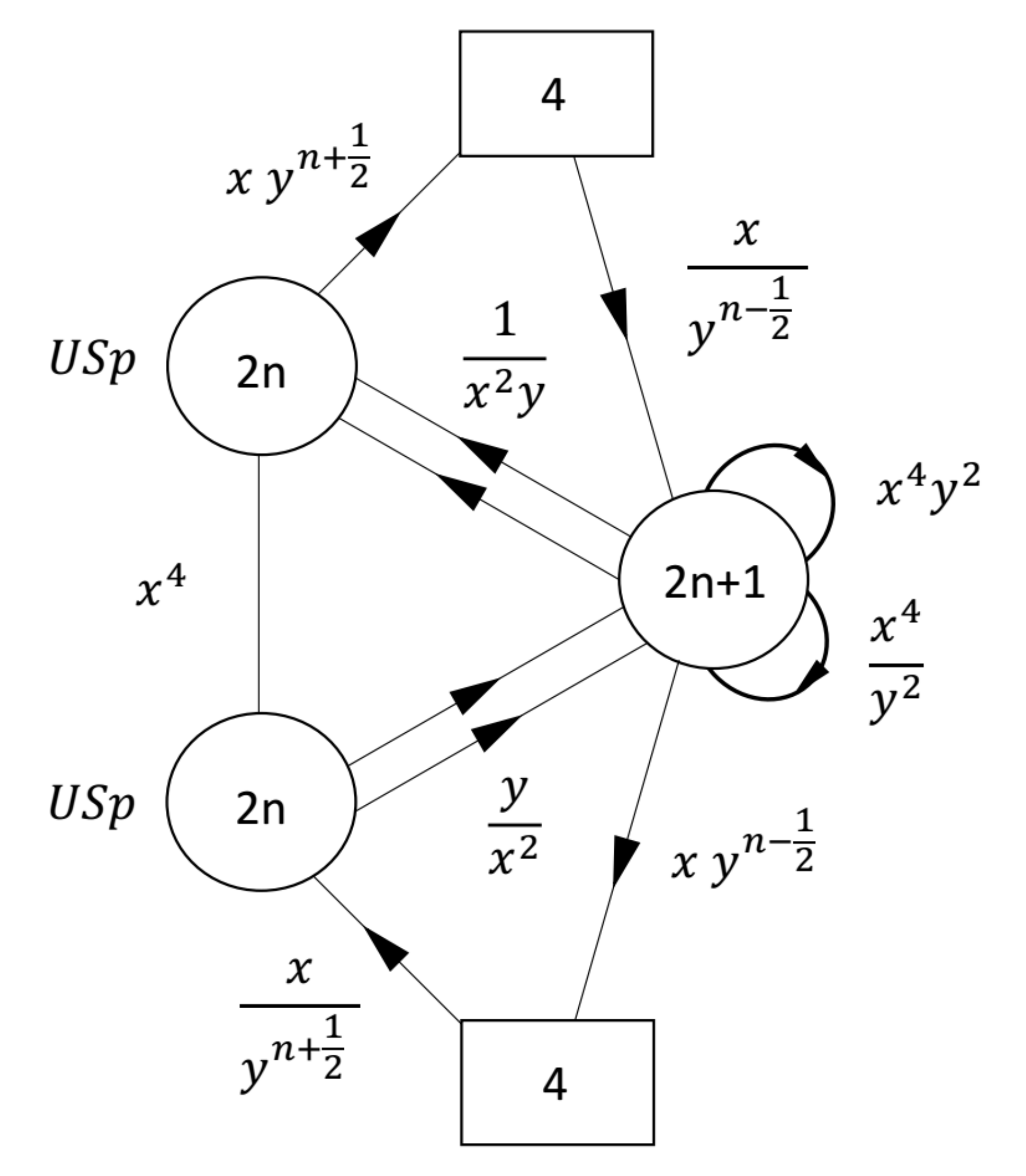} 
\caption{The proposed $4d$ quiver theory. There are cubic superpotentials along all triangles. As previously, we use arrows from the group to itself to denote antisymmetric chiral fields, where here one is in the antisymmetric representation, and the other in the conjugate antisymmetric representation. These two antisymmetric chirals are coupled to the appropriate bifundamentals via cubic superpotentials.}
\label{MDFQuiver1}
\end{figure}

First we recall the two $5d$ IR gauge theory descriptions of the $6d$ $SO(16)$ SCFT compactified on a circle. One is an $SU(2n+1)_0 + 2AS+8F$ gauge theory while the other is a $4F+USp(2n) \times USp(2n)+4F$ quiver gauge theory. Next we consider the $4d$ theory shown in figure \ref{MDFQuiver1}. This theory as one gauge theory description on one side and the other one on the other side, which are in turn connected via bifundamental fields. This fits the $5d$ motivated expectation of a theory extrapolating between the two descriptions. As we shall now show this theory has various features that are on one side quite suggestive of a $6d$ connection, but on the other side appear inconsistent with a direct compactification.  

First we note that the theory is free: the one loop $\beta$ function for all gauge groups vanishes and all superpotetials are cubic. This already makes it quite special. At the level of global symmetries it has two $SU(4)\times SU(4)$ non-abelian symmetry groups as well as two anomaly free $U(1)$ symmetry groups, where the charges of all the fields are illustrated in figure \ref{MDFQuiver1} using fugacities. The theory also has an anomaly free $U(1)_R$ symmetry where the $4$ $SU\times USp$ bifundamentals have R-charge $0$, the antisymmetrics and $USp \times USp$ bifundamental have R-charge $2$, while the rest of the fields have R-charge $1$. This R-symmetry has the special property that $Tr(R) = Tr(R^3)=0$, which is the result one finds for the Cartan of $SU(2)_R$ from integrating the $6d$ anomaly polynomial on the torus. Thus, it has the right properties to be the natural $6d$ R-symmetry that comes from the Cartan of $SU(2)_R$ in $6d$. The superconformal R-symetry, under which all fields have the free R-charge, is a mixture of this symmetry with $U(1)_x$, but does not involve $U(1)_y$.

These properties seem to hint on a $6d$ origin of the theory. However, there are several issues with trying to connect the theory to a direct compactification of the $6d$ $SO(16)$ SCFT. One problem is that the rank of the global symmetry seems to be too small. Particularly the $4d$ theory has rank $8$ global symmetry while the $6d$ SCFT has rank $10$ global symmetry. Yet, one may hope that fractional fluxes are the cause for the loss of some of the global symmetry.

Another issue is the R-charge assignment of the fields. Generically in $6d$ compactification, the $5d$ bulk fields have R-charge $1$, which usually includes all fields that appear in the $5d$ description. This can be seen from the $5d$ boundary condition viewpoint as these are associated with chiral fields in hypermultiplets of the $5d$ gauge theory that were given Nuemann boundary conditions, and these are part of a doublet of $SU(2)_R$. However, here the antisymmetrics and $USp \times USp$ bifundamental both have R-charge $2$ even though they are part of the $5d$ description. 

Returning to the issue of the global symmetry, it is instructive to try to see if the symmetry can be enhanced to a larger group. We have already noted that there is no mixing with $U(1)_y$ and so it is possible that it enhances to a larger group. Here we shall do a crude general analysis, reserving a more careful exact analysis in special cases for later. So let us consider the BPS objects we can build. The lowest dimension ones will be made out of only two fields and we can write down their contribution to the index by inspection. First we have various invariants that are charged only under $U(1)_x$, particularly the baryon made from the $USp \times USp$ bifundamental and the invariant made from the two antisymmetrics. Next we have several fields charged also under the non-abelian symmetries. These are the mesons of the two $USp$ groups and of the $SU$ group. These contribute fields with charges: $x^2 y^{2n+1} \chi[\bold{6},\bold{1}] + \frac{x^2}{y^{2n+1}} \chi[\bold{1},\bold{6}] + x^2\chi[\bold{\bar{4}},\bold{4}]$. We note that these can be combined into the $x^2 \chi[\bold{28}]$ of $U(1) \times SU(8)$ where $U(1)_y \times SU(4)\times SU(4)$ is embedded in $SU(8)$ as: $\bold{8}\rightarrow y^{n+\frac{1}{2}} \chi[\bold{\bar{4}},\bold{1}]+\frac{1}{y^{n+\frac{1}{2}}} \chi[\bold{1},\bold{4}]$.   
 
We are now in a position to try and connect this theory to a $6d$ compactification. Our claim is that this theory results from a mass deformation of the theory one gets when compactifying the $6d$ $SO(16)$ SCFT with flux $\frac{1}{2}$ breaking $SO(16)\rightarrow U(1)\times SU(8)$. Here the $6d$ $SU(2)_E\times SU(2)_F\times SO(16)$ global symmetry is broken to $U(1)_x\times SU(8)$ where $SU(2)_E$ is broken due to the fractional flux, $SU(2)_F$ is broken due to the mass deformation and $SO(16)$ is broken by the flux.

Our main evidence to backup this claim is that the anomalies can be matched. Let's start with the anomalies for the gauge theory in figure \ref{MDFQuiver1}. These can be conveniently packaged into an anomaly polynomial $6$ form, that for the case at hand reads:

\bea
I^{GT}_6 & = & \frac{4(48n^2+12n+1)}{3} C_1 (U(1)_x)^3 + 32 n^2 C_1 (U(1)_x)^2 C_1 (R) - 4n C_1 (U(1)_x) C_1 (R)^2 \label{GTAP} \\ \nonumber & + & (4n+1)(2n+1)^2 C_1 (U(1)_x) C_1 (U(1)_y)^2 - \frac{(3n+1)}{3} p_1 (T) C_1 (U(1)_x) \\ \nonumber & - & \frac{1}{2} (C_3 (SU(4)_1)_{\bold{4}}-C_3 (SU(4)_2)_{\bold{4}}) - (4n+1) C_1 (U(1)_x) (C_2 (SU(4)_1)_{\bold{4}}+C_2 (SU(4)_2)_{\bold{4}}) \\ \nonumber & - & (n+\frac{1}{2}) C_1 (U(1)_y) (C_2 (SU(4)_1)_{\bold{4}}-C_2 (SU(4)_2)_{\bold{4}})
\eea

As we pointed out earlier the spectrum shows some signs that it might have an enhanced $SU(8)$ symmetry. If this is indeed true then it must be possible to cast the anomaly polynomial in term of just the $SU(8)$ anomalies. Using the decomposition we find that:

\bea
& & C_2 (SU(8))_{\bold{8}} = C_2 (SU(4)_1)_{\bold{4}}+C_2 (SU(4)_2)_{\bold{4}} - (2n+1)^2 C_1 (U(1)_y)^2, \\ \nonumber
& & C_3 (SU(8))_{\bold{8}} = C_3 (SU(4)_2)_{\bold{4}}-C_3 (SU(4)_1)_{\bold{4}} - (2n+1) C_1 (U(1)_y) (C_2 (SU(4)_1)_{\bold{4}}-C_2 (SU(4)_2)_{\bold{4}}) .
\eea 

Using this we see that (\ref{GTAP}) can indeed be written as: 

\bea
I^{GT}_6 & = & \frac{4(48n^2+12n+1)}{3} C_1 (U(1)_x)^3 + 32 n^2 C_1 (U(1)_x)^2 C_1 (R) - 4n C_1 (U(1)_x) C_1 (R)^2 \nonumber \\ & - & \frac{(3n+1)}{3} p_1 (T) C_1 (U(1)_x) - (4n+1) C_1 (U(1)_x) C_2 (SU(8))_{\bold{8}} +\frac{1}{2} C_3 (SU(8))_{\bold{8}} \label{GTAPe}
\eea

Already this is a non-trivial consistency check on the $SU(8)$ enhancement.

Next we consider the $6d$ side, where we reduce the $SO(16)$ SCFT with flux $\frac{1}{2}$ breaking $SO(16)\rightarrow U(1)\times SU(8)$. Here we normalize the $U(1)$ such that $\bold{16} \rightarrow \bold{8}^1+\bold{\bar{8}}^{-1}$. There are two distinct such breaking differing by how the two $SO(16)$ spinors decompose. Since the matter spectrum of the $SO(16)$ SCFT appears to be asymmetric with respect to $SO(16)$ outer automorphisms, this should in principle be distinguishable. Here we consider the embedding where the $\bold{128}$ of $SO(16)$, where we use the same conventions as in section $2$, contains an $SU(8)$ singlet. With this choice one can see that the flux $\frac{1}{2}$ can be made consistent by incorporating a central flux in either $SU(2)_E$ or $SU(8)$, where here we shall choose the former case.

So next we want to perform the integration of the anomaly polynomial over the Riemann surface. But first we need to decompose the various characteristic classes. Performing the decomposition we find:

\bea
& & C_2 (SO(16))_{\bold{16}} = 2C_2 (SU(8))_{\bold{8}} - 8 C_1 (U(1))^2, \\ \nonumber
& & C_4 (SO(16))_{\bold{16}} = C^2_2 (SU(8))_{\bold{8}} + 28 C_1 (U(1))^4 - 10 C_1 (U(1))^2 C_2 (SU(8))_{\bold{8}} - 6 C_1 (U(1)) C_3 (SU(8))_{\bold{8}} \\ \nonumber & & + 2 C_4 (SU(8))_{\bold{8}}.
\eea  

Next we take $C_2(SU(2)_E)_{\bold{2}} = 0, C_1 (U(1)) = \frac{1}{2} t + C_1 (U(1)_x)$, where the latter inserts the flux into the $U(1)$ while the former takes into account the fact that $SU(2)_E$ is completely broken due to center fluxes. Performing the calculation we find:

\bea
I^{6d}_6 & = & \frac{4(12n+1)}{3} C_1 (U(1)_x)^3 - 4n(n+1) C_1 (U(1)_x) C_1 (R)^2 - \frac{(3n+1)}{3} p_1 (T) C_1 (U(1)_x) \nonumber \\ & - & 4n^2 C_2(SU(2)_F)_{\bold{2}} C_1 (U(1)_x)  - (4n+1) C_1 (U(1)_x) C_2 (SU(8))_{\bold{8}} +\frac{1}{2} C_3 (SU(8))_{\bold{8}} . \nonumber \\ \label{6dAP}
\eea 

Finally we can decompose $SU(2)_F$ to its Cartan $U(1)_F$ under which $C_2(SU(2)_F)_{\bold{2}} = - C_1 (U(1)_F)^2$. If we now take $C_1 (U(1)_F) = C_1 (R) + 4 C_1 (U(1)_x)$ then equations (\ref{6dAP}) and (\ref{GTAPe}) match. This suggests that the theory in figure \ref{MDFQuiver1} is a deformation of a torus reduction of the $SO(16)$ SCFT with flux $\frac{1}{2}$ breaking $SO(16)$ to $U(1)\times SU(8)$. The deformation has the effect of breaking $SU(2)_F$, locking its Cartan to a combination of $U(1)_x$ and $U(1)_R$. We shall next consider a specific example, the one with $n=1$, where we can actually show this explicitly and identify the deformation as a mass deformation. A by product of that is an interesting dual description of the theory in figure \ref{MDFQuiver1}.  

Consider the case of $n=1$. In this case the $6d$ SCFT is the minimal $(D_5,D_5)$ conformal matter and the $6d$ global symmetry enhances to $SO(20)$. However the flux breaks it to $U(1)\times SU(8)\times SO(4)$ so we do not expect any special enhancement of symmetry in the $4d$ theory in figure \ref{MDFQuiver1}.

We can evaluate the index of this theory in this case in order to check the enhancement to $SU(8)$. Performing the calculation up to order $p q$, we indeed find the index can be arranged into $SU(8)$ characters where it reads:

\be
I^{n=1}_{4d} = 1 + (p q)^{\frac{2}{3}} (2x^8 + x^5 \chi[\bold{8}] + x^2 \chi[\bold{28}]) + p q (\frac{2}{x^3}\chi[\bold{8}]+2+2x^3\chi[\bold{\bar{8}}]+x^6\chi[\bold{\bar{28}}]) + ... \label{IndSp}
\ee

Here as expected: $\chi[\bold{8}]= y^{\frac{3}{2}} \chi[\bold{\bar{4}},\bold{1}]+\frac{1}{y^{\frac{3}{2}}} \chi[\bold{1},\bold{4}]$. 

\begin{figure}
\center
\includegraphics[width=0.75\textwidth]{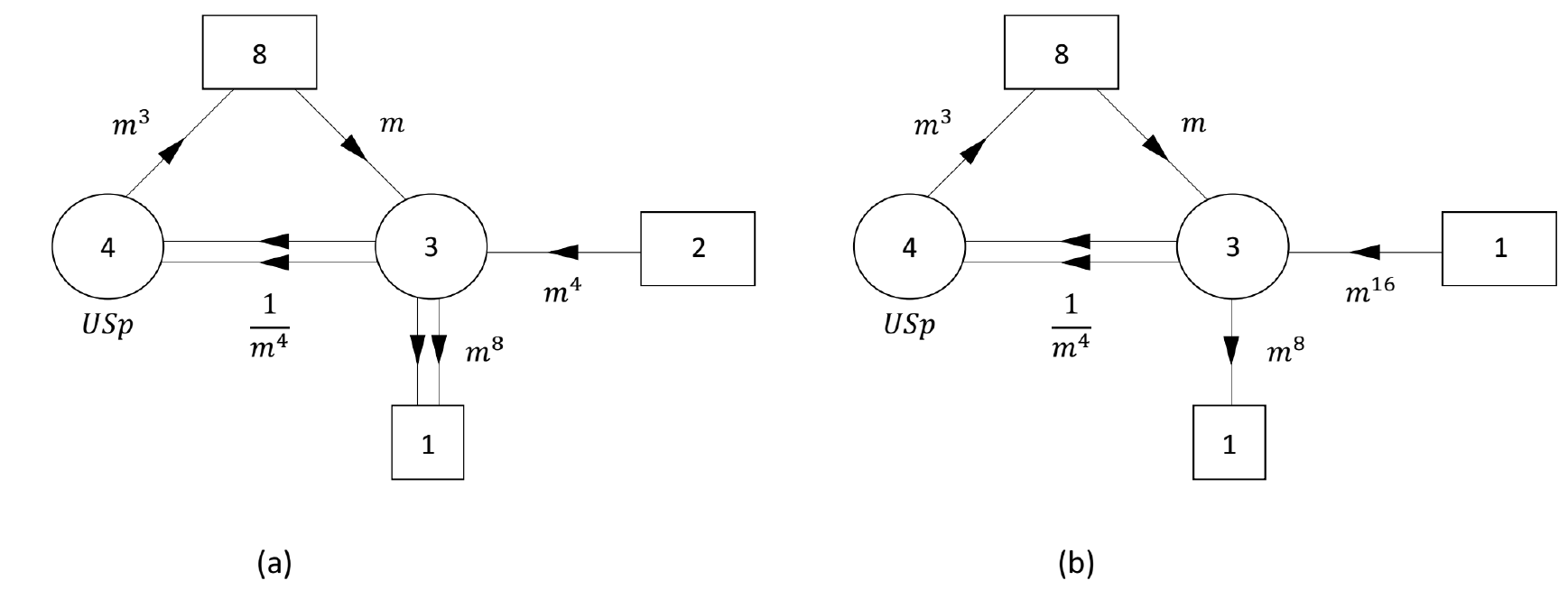} 
\caption{(a) The $4d$ theory associated in \cite{KRVZ1} with the compactification of the minimal $(D_5,D_5)$ conformal matter on a torus with flux $\frac{1}{2}$ preserving $U(1)\times SU(8)\times SO(4)$. (b) The $4d$ theory we get after a mass deformation giving mass to two flavors of the $SU(3)$ group. Here in addition to the cubic superpotentials there is a quintic superpotentials involving the $m^{16}$ flavor and $4$ bifundamentals.}
\label{DDCMQuivers}
\end{figure}

The compactification of the minimal $(D,D)$ conformal matter on a torus with fluxes was studied in \cite{KRVZ1}, and we can thus perform a direct comparison with results obtained there. We start with the theory \cite{KRVZ1} associated with a compactification of the minimal $(D_5,D_5)$ conformal matter on a torus with flux $\frac{1}{2}$ preserving $U(1)\times SU(8)\times SO(4)$. This theory can be generated by gluing a tube with that flux to itself. The resulting $4d$ theory is shown in figure \ref{DDCMQuivers} (a). We can evaluate the index for this theory finding:

\bea
I^{DDCM}_{4d} & = & 1 + (p q)^{\frac{2}{3}} m^6 \chi[\bold{28}] + (p q)^{\frac{5}{6}} (h+\frac{1}{h}) (2m^{12} + m^3 \chi[\bold{8}]) + p q (3m^9\chi[\bold{\bar{8}}] + \frac{1}{m^9}\chi[\bold{8}]) \\ \nonumber & + & (p q)^{\frac{2}{3}}(p+q) m^6 \chi[\bold{28}] + (p q)^{\frac{7}{6}} (h+\frac{1}{h}) (m^{6}\chi[\bold{\bar{28}}] - \frac{1}{m^3} \chi[\bold{\bar{8}}]) \\ \nonumber & + & (p q)^{\frac{5}{6}}(p+q) (h+\frac{1}{h}) (2m^{12} + m^3 \chi[\bold{8}]) + (p q)^{\frac{4}{3}} (\frac{1}{m^{24}} - \frac{1}{m^{15}}\chi[\bold{\bar{8}}] - \frac{1}{m^{6}}\chi[\bold{\bar{28}}]+ m^3 \chi[\bold{\bar{56}}] \\ \nonumber & + & m^{12} (\chi[\bold{336}]+\chi[\bold{70}])) + ...
\eea 

Here $U(1)_m$ is the flux $U(1)$ normalized such that $\bold{20}_{SO(20)} \rightarrow (\bold{2},\bold{2},\bold{1})_{SU(2)\times SU(2)\times SU(8)} + m^3 (\bold{1},\bold{1},\bold{8})_{SU(2)\times SU(2)\times SU(8)} + \frac{1}{m^3} (\bold{1},\bold{1},\bar{\bold{8}})_{SU(2)\times SU(2)\times SU(8)}$, $h$ is the fugacity for $SU(2)_F$ in our notation, and the second $SU(2)$ was broken by the flux. Note that in evaluating the index we have used the R-symmetry: $U(1)_{6d} - \frac{1}{9} U(1)_m$ which while not the superconformal one, is quite close to it. The superconformal R-symmetry is actually $U(1)_{6d} - \frac{\sqrt{22}}{9\sqrt{13}} U(1)_m$. Comparing this index with (\ref{IndSp}) we see that they match if we take $h = \frac{m^{12}}{(p q)^{\frac{1}{6}}}, m^3 = x$. This implies that the mass deformation needs to equate $U(1)_F =  12 U(1)_m - \frac{1}{3} U(1)_R = U(1)^{6d}_R + 4 U(1)_x$ as expected from anomalies.

It is interesting to identify the exact deformation one must do. The identification we found would be naturally generated if a term with charges $\frac{m^{12} (p q)^{\frac{5}{6}}}{h}$ was added to the superpotential. This term indeed appears in the index and so the required relevant deformation indeed exists. We can next identify what is the operator in terms of the fields of the $4d$ theory in figure \ref{DDCMQuivers} (a). From the charges it is clear that it is the mass term involving one of the two $SU(2)$ doublets and one of the two $m^8$ flavors of the $SU(3)$ gauge group. However, since this is just a flavor mass term, we can perform the deformation directly on the Lagrangian by integrating out the two fields. This leads to the field theory shown in figure \ref{DDCMQuivers} (b). Besides removing the massive pair of chirals the integrating out also changed the charges slightly and induces a quintic superpotentials involving the $m^{16}$ flavor and $4$ bifundamentals. The charges of the fields under $U(1)_m$ are given in the figure, and those under $U(1)^{6d}_R$ are just $0$ for the bifundamentals, $2$ for the $SU(8)$ neutral $SU(3)$ flavors, and $1$ for all the others. 

We are now lead to the following conclusion: the theories shown in \ref{DDCMQuivers} (b) and \ref{MDFQuiver1}, for $n=1$, are dual. As we have calculated previously both the anomalies and index matches, at least to the order we computed. In the theory in figure \ref{DDCMQuivers} (b), the $SU$ group is conformal, while the $USp$ group is asymptotically free so we expect the theory to flow to an interacting fixed point. We do note that there is a quintic superpotential, however removing it does not generate a new symmetry so it does not appear to be important for the flow. The claim then is that the resulting theory is an SCFT with a conformal manifold that has a region with a weakly coupled description given by the theory in figure \ref{MDFQuiver1}. The price payed for the pleasure of such a description is that the $SU(8)$ symmetry that is manifest in the original description is broken by the marginal deformations.  

\begin{figure}
\center
\includegraphics[width=0.75\textwidth]{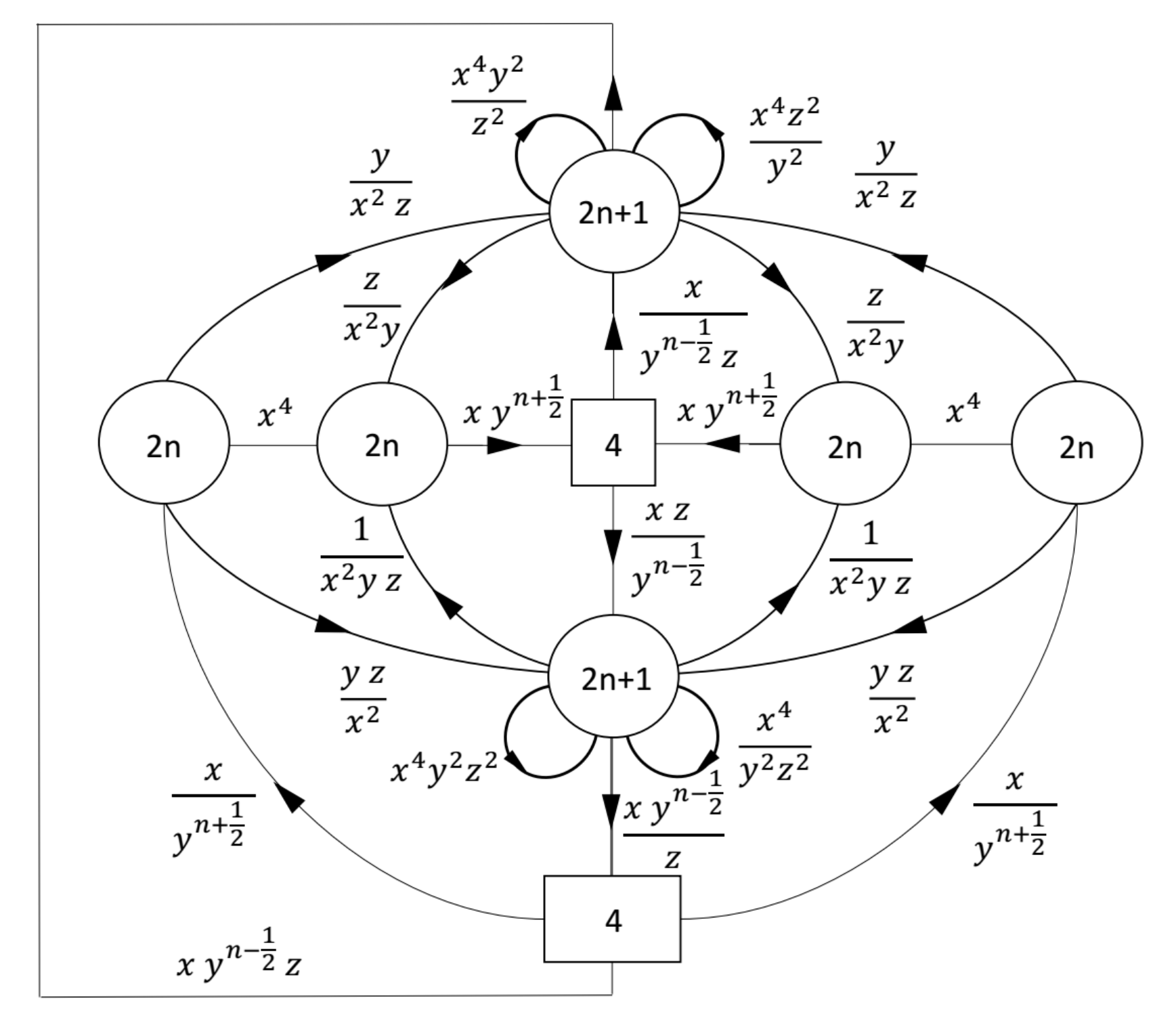} 
\caption{The $4d$ quiver theory associated to compactification and mass deformation, but now with flux $1$. The $2n$ groups are of type $USp$ while the $2n+1$ ones are of type $SU$.}
\label{NMDFQuiver1}
\end{figure}

It is straightforward to generalize all the above discussion to higher value of flux. One simply multiplies the matter content and connects them with bifundamentals such that the pair of bifundamentals in the theory of figure \ref{MDFQuiver1} is replaced with a circle of bifundamental that connects the $SU(2n+1)$ groups of all the copies with the $USp(2n)$ groups of the same and neighboring copy. The resulting theory for the case of flux $1$ is shown in figure \ref{NMDFQuiver1}. One interesting aspect here is that when the flux is integer we expect there to be an additional anomaly free $U(1)$ that should enhance to $SU(2)$. Indeed the theory in figure \ref{NMDFQuiver1} has an additional non-anomalous $U(1)$ as expected. 

We can again compare anomalies now involving this $U(1)_z$ symmetry. For instance in the theory in figure \ref{NMDFQuiver1}, the only term related to this symmetry in the anomaly polynomial is $I^{GT}_6 \supset 8(2n+1)^2 C_1 (U(1)_x) C_1 (U(1)_z)^2$. This matches the results from the $6d$ anomaly polynomial if we identify $\bold{2}_{SU(2)_E} = z^{2n+1} + \frac{1}{z^{2n+1}}$.

For the $n=1$ case, we can again study the theory before the mass deformation. This should give us a dual description of this theory.   

\subsection{$SU(2)\times E_7$ SCFT}

\begin{figure}
\center
\includegraphics[width=0.5\textwidth]{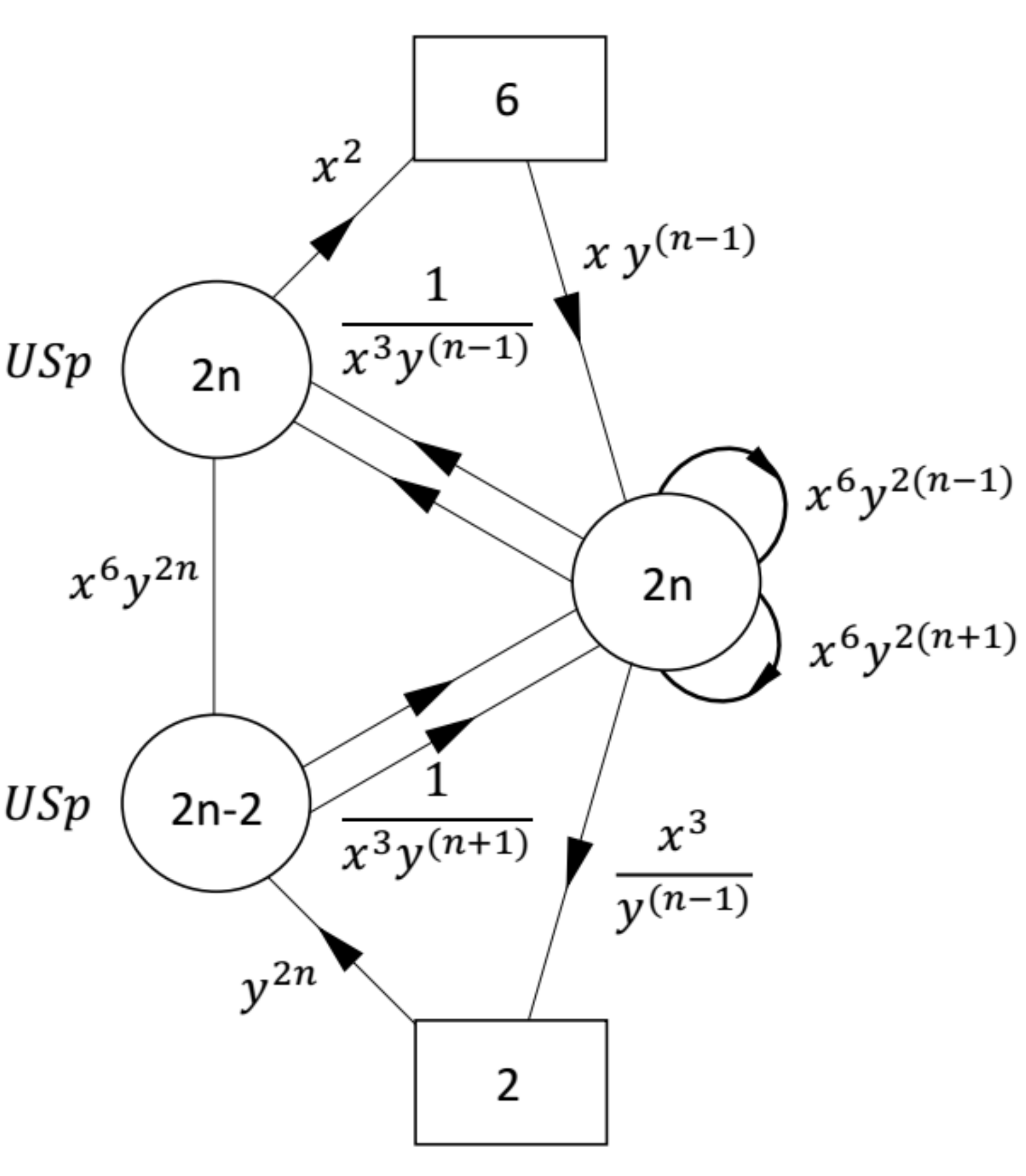} 
\caption{The proposed $4d$ quiver theory. There are cubic superpotentials along all triangles. As previously, we use arrows from the group to itself to denote antisymmetric chiral fields, where here one is in the antisymmetric representation, and the other in the conjugate antisymmetric representation. These two antisymmetric chirals are coupled to the appropriate bifundamentals via cubic superpotentials.}
\label{MDFQuiver2}
\end{figure}

For the $SU(2)\times E_7$ SCFT the two $5d$ IR gauge theory descriptions are an $SU(2n)_0 + 2AS+8F$ gauge theory and a $6F+USp(2n) \times USp(2n-2)+2F$ quiver gauge theory. Next we consider the $4d$ theory shown in figure \ref{MDFQuiver2}. This theory has one gauge theory description on one side and the other one on the other side, which are in turn connected via bifundamental fields. This fits the $5d$ motivated expectation of a theory extrapolating between the two descriptions. Similarly to the previously discussed theory, it is related to a compactification of the $SU(2)\times E_7$ SCFT on a torus with flux via a mass deformation breaking $SU(2)_F$. One notable difference between the two cases is that the flux used here is somewhat more complicated, which in turn leads to the $4d$ dynamics being more involved.  

 At the level of global symmetries we now have an $SU(2)\times SU(6)$ non-abelian symmetry group, as well as two anomaly free $U(1)$ symmetry groups, where the charges of all the fields are illustrated in figure \ref{MDFQuiver2} using fugacities. The theory also has an anomaly free $U(1)_R$ symmetry where the $4$ $SU\times USp$ bifundamentals have R-charge $0$, the antisymmetrics and $USp \times USp$ bifundamental have R-charge $2$, while the rest of the fields have R-charge $1$, exactly as in the previous case. Similarly, this R-symmetry has the special property that $Tr(R) = Tr(R^3)=0$, as expected from the $6d$ interpretation. Unlike the previous case, however, this theory is not free. Here the $USp(2n)$ group is asymptotically free, being one flavor short of conformality, while the $USp(2n-2)$ group is IR free, having one flavor in excess of conformality. The $SU(2n)$ group is free though. One can use a maximization to study this theory. We find that in general both $U(1)$ groups mix with $U(1)^{6d}_R$, and that the mixing is generically an unappealing mathematical function of $n$. We shall not dwell on the dynamics of this theory any further.

Instead we now seek to compute anomalies and compare them against the $6d$ interpretation. Particularly, our claim is that this $4d$ theory is generated by an $SU(2)_F$ breaking mass deformation of a theory that is a torus compactification of the $SU(2)\times E_7$ SCFT with flux $\frac{1}{2}$ in the Cartan of $SU(2)_I$, and flux $\frac{1}{2}$ in $E_7$ preserving $U(1)_{E_7}\times E_6$. Particularly, adopting a normalization convention where the minimal charge is $1$, we find that $U(1)_{E_7}= 2 U(1)_x$, and $U(1)_{SU(2)_I}= 2n U(1)_y$. Again for the flux to be consistent, it must also be accompanied by central fluxes in $SU(2)_E$ which in turn break this symmetry.

One expectation from this relation is that the anomalies be consistent with enhancement to $E_6$ and that the spectrum could be arranged in $E_6$ characters, where we expect that the $SU(2)\times SU(6)$ global symmetry should be realized as the maximal subgroup of $E_6$. We shall shortly show that the anomalies are consistent with this, but first we shall see that the basic matter spectrum forms the correct characters. Again for this we consider the basic gauge invariants we can build from the smallest number of fields charged under the non-abelian flavor symmetries. From just two fields, we have the gauge invariants consisting of the $SU(6)$ charged mesons of the $USp(2n)$ group, as well as the mesons of the $SU(2n)$ group. These contribute to the index the term: $p q x^4 (\chi[\bold{15}]_{SU(6)} + \chi[\bar{\bold{6}}]_{SU(6)} \chi[\bold{2}]_{SU(2)})$. From three fields, we have the gauge invariants consisting of the $USp(2n)$ and $USp(2n-2)$ charged flavors, connected via the $USp\times USp$ bifundamental, as well as the invariant made from two $SU(6)$ charged $SU(2n)$ flavors and one antisymmetric chiral. These contribute to the index the term: $p^2 q^2 x^8 y^{4n} (\chi[\bar{\bold{15}}]_{SU(6)} + \chi[\bold{6}]_{SU(6)} \chi[\bold{2}]_{SU(2)})$. Both can be cast into $E_6$ characters as, under its $SU(2)\times SU(6)$ maximal subgroup,we have that $\bold{27}_{E_6} \rightarrow \chi[\bold{15}]_{SU(6)} + \chi[\bar{\bold{6}}]_{SU(6)} \chi[\bold{2}]_{SU(2)}$ and likewise for the complex conjugate.

Nevertheless, our main evidence in favor of this claim is that the anomalies can be matched. Let us start with the anomalies for the gauge theory in figure \ref{MDFQuiver2}. These can be conveniently packaged into an anomaly polynomial $6$ form, that for the case at hand reads:

\bea
I^{GT}_6 & = & 72n(3n-2) C_1 (U(1)_x)^3 + 72 n(n-1) C_1 (U(1)_x)^2 C_1 (R) - 6n C_1 (U(1)_x) C_1 (R)^2 \label{GTAP2} \\ \nonumber & + & 24n^2 (3n^2 -2n + 1) C_1 (U(1)_x) C_1 (U(1)_y)^2 + 24n (9n^2 -8n + 2) C_1 (U(1)_x)^2 C_1 (U(1)_y) \\ \nonumber & + & 24n (2n^2 -2n + 1) C_1 (U(1)_x) C_1 (U(1)_y) C_1 (R) + \frac{8n^3(3n^2+1)}{3} C_1 (U(1)_y)^3 \\ \nonumber & + & 8 n^2(n^2 -n +1) C_1 (U(1)_y)^2 C_1 (R) - 2n(n-2) C_1 (U(1)_y) C_1 (R)^2 - \frac{3n}{2} p_1 (T) C_1 (U(1)_x) \\ \nonumber & - & \frac{n(3n-2)}{6} p_1 (T) C_1 (U(1)_y) - 6n C_1 (U(1)_x) (C_2 (SU(6))_{\bold{6}}+C_2 (SU(2))_{\bold{2}}) \\ \nonumber & - & 2n(n-1) C_1 (U(1)_y) (C_2 (SU(6))_{\bold{6}}+C_2 (SU(2))_{\bold{2}}) .
\eea

We expect that the $SU(2)\times SU(6)$ dependent anomalies can be written in terms of just $E_6$ anomalies. Using the decomposition we find that:

\be
C_2 (E_6)_{\bold{27}} = 6 C_2 (SU(6))_{\bold{6}} + 6 C_2 (SU(2))_{\bold{2}}.
\ee 

Using this we see that (\ref{GTAP2}) can indeed be written as: 

\bea
I^{GT}_6 & = & 72n(3n-2) C_1 (U(1)_x)^3 + 72 n(n-1) C_1 (U(1)_x)^2 C_1 (R) - 6n C_1 (U(1)_x) C_1 (R)^2 \label{GTAPe2} \\ \nonumber & + & 24n^2 (3n^2 -2n + 1) C_1 (U(1)_x) C_1 (U(1)_y)^2 + 24n (9n^2 -8n + 2) C_1 (U(1)_x)^2 C_1 (U(1)_y) \\ \nonumber & + & 24n (2n^2 -2n + 1) C_1 (U(1)_x) C_1 (U(1)_y) C_1 (R) + \frac{8n^3(3n^2+1)}{3} C_1 (U(1)_y)^3 \\ \nonumber & + & 8 n^2(n^2 -n +1) C_1 (U(1)_y)^2 C_1 (R) - 2n(n-2) C_1 (U(1)_y) C_1 (R)^2 - \frac{3n}{2} p_1 (T) C_1 (U(1)_x) \\ \nonumber & - & \frac{n(3n-2)}{6} p_1 (T) C_1 (U(1)_y) - n C_1 (U(1)_x) C_2 (E_6)_{\bold{27}} - \frac{n(n-1)}{3} C_1 (U(1)_y) C_2 (E_6)_{\bold{27}},
\eea
so the anomalies are consistent with $E_6$.

Next we consider the $6d$ side, where we reduce the $SU(2)\times E_7$ SCFT with flux $\frac{1}{2}$ breaking $E_7\rightarrow U(1)_{E_7}\times E_6$, and also flux $\frac{1}{2}$ in the Cartan of $SU(2)_I$. Here we normalize the $U(1)$ groups such that the minimal charge is $1$. As previously noted, consistency requires that this flux be accompanied with central flux for $SU(2)_E$, casing it to be completely broken in the $4d$ theory. 

So next we want to perform the integration of the anomaly polynomial over the Riemann surface. But first we need to decompose the various characteristic classes. Performing the decomposition we find:

\bea
& & C_2 (SU(2)_I)_{\bold{2}} = - C_1 (U(1)_{SU(2)_I})^2, \\ \nonumber
& & C_2 (E_7)_{\bold{56}} = 2C_2 (E_6)_{\bold{27}} - 36 C_1 (U(1)_{E_7})^2.
\eea  

Next we take $C_2(SU(2)_E)_{\bold{2}} = 0, C_1 (U(1)_{E_7}) = \frac{1}{2} t + 2 C_1 (U(1)_x), C_1 (U(1)_{SU(2)_I}) = \frac{1}{2} t + 2n C_1 (U(1)_y)$, where the latter two insert the flux in the appropriate $U(1)$ groups while the former takes into account the fact that $SU(2)_E$ is completely broken due to center fluxes. Performing the calculation we find:

\bea
I^{6d}_6 & = & 72 n C_1 (U(1)_x)^3 + 24 n(n-1) C_1 (U(1)_x)^2 C_1 (U(1)_y) + 24 n^2(n-1) C_1 (U(1)_y)^2 C_1 (U(1)_x) \nonumber \\ & + & \frac{8n^2(3n-2)}{3} C_1 (U(1)_y)^3 - 6n^2 C_1 (U(1)_x) C_1 (R)^2 - 2n(n^2-1) C_1 (U(1)_y) C_1 (R)^2 \nonumber \\ & - & \frac{3 n}{2} p_1 (T) C_1 (U(1)_x) - \frac{(3n-2)}{6} p_1 (T) C_1 (U(1)_y) - n C_1 (U(1)_x) C_2 (E_6)_{\bold{27}} + C_1 (U(1)_y) C_2 (E_6)_{\bold{27}} \nonumber \\ & - & 6n(n-1) C_2(SU(2)_F)_{\bold{2}} C_1 (U(1)_x) + 6(n^2-n+1) C_2(SU(2)_F)_{\bold{2}} C_1 (U(1)_y) . \label{6dAP2}
\eea 

Finally we can decompose $SU(2)_F$ to its Cartan $U(1)_F$ under which $C_2(SU(2)_F)_{\bold{2}} = - C_1 (U(1)_F)^2$. If we now take $C_1 (U(1)_F) = C_1 (R) + 6 C_1 (U(1)_x) + 2n C_1 (U(1)_y)$ then equations (\ref{6dAP2}) and (\ref{GTAPe2}) match.

 We can again consider theories corresponding to higher values of flux by adjoining two copies of the theory in figure \ref{MDFQuiver2} so as to form a circle, similarly to the theory in figure \ref{NMDFQuiver1}. We expect this theory to be related to a similar compactification and mass deformation but with flux $1$, and similarly for higher values of flux. However, we shall not check this explicitly here. 

\section{Conclusions}

Here we have explored the torus compactifications of $Z_2$ orbifold of E-string theories with non-trivial action on the $E_8$ group. We have conjectured $4d$ theories corresponding to a tube with flux $\frac{1}{2}$ in the $SU(2)$ global symmetry associated with the orbifold in these $6d$ SCFTs. We have further tested this by various methods, notably by comparing anomalies against those expected from $6d$. For the special case of the $SO(7)\times E_7$ conformal matter, this can be extended to more tubes that together cover a wide variety of possible fluxes in $SO(7)$. These $4d$ theories should exhibit interesting phenomena of symmetry enhancement which is ultimately attributed to the symmetry inherited from the $6d$ SCFT. This was tested using anomalies, and in specifically simple cases, also the superconformal index.

We have also studied a class of theories related to $6d$ torus compactifications by mass deformations. We supported this claim by matching anomalies, and in the special case of the $(D_5,D_5)$ conformal matter also using the superconformal index compared against the mass deformation of the conjectured direct reduction. These, while not direct $6d$ compactifications, still retain some of the miraculous properties of $6d$ compactifications. Particularly, they still retain some of the $6d$ global symmetry which can lead to interesting symmetry enhancement phenomena. Also the general idea of pair of pants decomposition leading to $4d$ dualities still applies also to mass deformed theories. Indeed, the study of these theories for the case of the $(D_5,D_5)$ conformal matter lead us to an apparently new duality that can be ultimately linked to the different $5d$ descriptions of the $6d$ $(D_5,D_5)$ conformal matter on a circle. Similar dualities were also proposed in \cite{KRVZ1,KRVZ2}.

There are various future directions one might consider. One is to consider the generalizations to larger orbifold group. Here we have seen that there are Lagrangian theories that one can associate to a torus compactification with flux in the global symmetry related to the orbifold. It is interesting if this persists also to more general orbifolds. 

We can also consider attempting to realize more values of flux. Ultimately the flux values we have realized here are quite limited, and leave much to be desired. It is possible that considering more intricate systems, still retaining the structure expected from the $5d$ domain wall picture, can provide the other cases.

There are also several interesting issues in the $5d$ picture, particularly the relation between the $5d$ gauge theory and $6d$ SCFT. For example, it should be possible to determine the flux directly from the holonomies on the two sides. These then may give us a tool to understand which fluxes can be realized via a domain wall between $5d$ gauge theories, and so are expected to give Lagrangian theories in $4d$, and which ones may not.

\section*{Acknowledgments}
We thank Hee-Cheol Kim, Shlomo Razamat, and Cumrun Vafa for relevant discussions and for participation in early stages of this project. GZ is supported in part by World Premier International Research Center Initiative (WPI), MEXT, Japan.


\begin{thebibliography}{40}

\bibitem{Gai} 
  D.~Gaiotto,
  JHEP {\bf 1208}, 034 (2012)
  [arXiv:0904.2715 [hep-th]].

\bibitem{BTW} 
  F.~Benini, Y.~Tachikawa, and B.~Wecht,
	JHEP {\bf 1001}, 088 (2010)
  [arXiv:0909.1327 [hep-th]].

\bibitem{BBBW}
  I.~Bah, C.~Beem, N.~Bobev, and B.~Wecht,
	JHEP {\bf 1206}, 005 (2012)
  [arXiv:1203.0303 [hep-th]].

\bibitem{GR}
  D.~Gaiotto, S.~S.~Razamat,
  JHEP {\bf 1507}, 073 (2015)
  [arXiv:1503.05159 [hep-th]].

\bibitem{FHU}
  S.~Franco, H.~Hayashi, and A.~Uranga,
	Phys.\ Rev.\  D92:no.4, 045004 (2015)
  [arXiv:1504.05988 [hep-th]].

\bibitem{HM}
  A.~Hanany, K.~Maruyoshi,
  JHEP {\bf 1512}, 080 (2015)
  [arXiv:1505.05053 [hep-th]].

\bibitem{RVZ}
  S.~S.~Razamat, C.~Vafa, and G.~Zafrir,
  JHEP {\bf 1704}, 064 (2017) 
  [arXiv:1610.09178 [hep-th]].

\bibitem{BHMRTZ} 
  I.~Bah, A.~Hanany, K.~Maruyoshi, S.~S.~Razamat, Y.~Tachikawa, and G.~Zafrir,
	JHEP {\bf 1706}, 022 (2017)
  [arXiv:1702.04740 [hep-th]].

\bibitem{KRVZ} 
  H.~-C.~Kim, S.~S.~Razamat, C.~Vafa, and G.~Zafrir,
	Fortsch.Phys. 66 (2018) no.1, 1700074
  [arXiv:1709.02496 [hep-th]].
	
\bibitem{KRVZ1} 
  H.~-C.~Kim, S.~S.~Razamat, C.~Vafa, and G.~Zafrir,
	JHEP {\bf 1806}, 058 (2018)
  [arXiv:1802.00620 [hep-th]].

\bibitem{KRVZ2} 
  H.~-C.~Kim, S.~S.~Razamat, C.~Vafa, and G.~Zafrir,
  [arXiv:1806.07620 [hep-th]].
	
\bibitem{RZ} 
  S.~S.~Razamat, G.~Zafrir,
  [arXiv:1806.09196 [hep-th]].

\bibitem{Zaf2} 
  G.~Zafrir,
	JHEP {\bf 1512}, 157 (2015)
  arXiv:1509.02016 [hep-th].

\bibitem{HKLY} 
  H.~Hayashi, S.~Kim, K.~Lee, and F.~Yagi,
  [arXiv:1509.03300 [hep-th]].

\bibitem{HMRV}
  J.~J.~Heckman, D.~R.~Morrison, T.~Rudelius, and C.~Vafa,
	Fortsch.Phys. 63 (2015) 468-530
  [arXiv:1502.05405 [hep-th]].

\bibitem{MOTZ} 
  N.~Mekareeya, K.~Ohmori, Y.~Tachikawa, and G.~Zafrir,
	JHEP {\bf 1709}, 144 (2017)
  [arXiv:1707.04370 [hep-th]].

	
\bibitem{OSTY2} 
  K.~Ohmori, H.~Shimizu, Y.~Tachikawa, and K.~Yonekura,
	JHEP {\bf 1512}, 131 (2015)
  [arXiv:1508.00915 [hep-th]].

\bibitem{ZHTV}
  M.~Del Zotto, J.~J.~Heckman, A.~Tomasiello, and C.~Vafa,
  JHEP {\bf 1502}, 054 (2015) 
  [arXiv:1407.6359 [hep-th]].

\bibitem{HMa}
  A.~Hanany, N.~Mekareeya,
  JHEP {\bf 1807}, 098 (2018)
  [arXiv:1801.01129 [hep-th]].

\bibitem{OSTY1} 
  K.~Ohmori, H.~Shimizu, Y.~Tachikawa, and K.~Yonekura,
	JHEP {\bf 1507}, 014 (2015)
  [arXiv:1503.06217 [hep-th]].

\bibitem{DVX}
  M.~Del Zotto, C.~Vafa, and D.~Xie,
	JHEP {\bf 1511}, 123 (2015)
  [arXiv:1504.08348 [hep-th]].

\bibitem{OS} 
  K.~Ohmori, H.~Shimizu,
	JHEP {\bf 1603}, 024 (2016)
  [arXiv:1509.03195 [hep-th]].

\bibitem{GC}
  D.~Gaiotto, H.~-C.~Kim,
	JHEP {\bf 1701}, 019 (2017)
  [arXiv:1506.03871 [hep-th]].

\bibitem{BG}
  C.~Beem, A.~Gadde,
	JHEP {\bf 1404}, 036 (2014)
  [arXiv:1212.1467 [hep-th]].

\bibitem{ABMS}
  P.~Agarwal, I.~Bah, K.~Maruyoshi, and J.~Song, 
	JHEP {\bf 1503}, 049 (2015)
  [arXiv:1409.1908 [hep-th]].
	
\bibitem{OSTY} 
  K.~Ohmori, H.~Shimizu, Y.~Tachikawa, and K.~Yonekura,
	PTEP 2014 10, 103B07 (2014)
  [arXiv:1408.5572 [hep-th]].
	
	\bibitem{BRZ} 
  C.~Beem, S.~S.~Razamat, and G.~Zafrir,
	to appear.

\bibitem{Wit} 
  E.~Witten,
	JHEP {\bf 9802}, 006 (1998)
  [arXiv:9712028 [hep-th]].

\end{thebibliography}
\end{document}